\documentclass[paper]{elsarticle}
\usepackage{color}
\usepackage{multirow}
\usepackage{graphicx}
\usepackage{subfigure}
\usepackage{epsfig}
\usepackage{rotating}
\usepackage{amsmath, amssymb}
\usepackage{pdflscape}
\usepackage{bbm}
\usepackage{afterpage}
\usepackage{tikz}
\usetikzlibrary{shapes,arrows}

\newtheorem{remark}{Remark}[section]

\newtheorem{lemma}{\bf Lemma}

\usepackage{lineno,hyperref}
\modulolinenumbers[5]

\journal{Communications in Nonlinear Science and Numerical Simulation}
\date{}
\newenvironment{proof}{
\begin{trivlist}
\item[\hspace{\labelsep}{\bf\noindent Proof. }] }
{\par\hfill\end{trivlist}\par}

\begin{document}
\begin{frontmatter}
	\title{\bf
		Study of a general growth model}
	\author[myfirstmainaddress]{G. { Albano}}
	\author[mymainaddress]{\ V. { Giorno}}
	\author[mysecondmainaddress]{\ P. { Rom\'an-Rom\'an}}
	\author[mysecondmainaddress]{\ F. { Torres-Ruiz}}
	\address{Email: pialbano@unisa.it \ -- \ ORCID: 0000-0002-2317-0331\\
		Email: giorno@unisa.it \ -- \ ORCID: 0000-0001-6474-3952\\
		Email: proman@ugr.es \ -- \ ORCID: 0000-0001-7752-8290\\
		Email: fdeasis@ugr.es \ -- \ ORCID:  0000-0001-6254-2209}
	\address[myfirstmainaddress]{Dipartimento di Studi Politici e Sociali, Universit\`a degli Studi di Salerno, Via Giovanni Paolo II n.\ 132, I-84084 Fisciano (SA), Italy}
	\address[mymainaddress]{Dipartimento di Informatica, Universit\`a degli Studi di Salerno, Via Giovanni Paolo II n.\ 132, I-84084 Fisciano (SA), Italy}
	\address[mysecondmainaddress]{Department of Statistics and Operations Research, Faculty of Sciences, University of Granada, 18071 Granada, Spain; Institute of Mathematics of the University of Granada (IMAG), Calle Ventanilla, 11, 18001, Granada, Spain }
	\begin{abstract}
		We discuss a general growth curve including several parameters, whose choice leads to a variety of models including the classical cases of Malthusian, Richards, Gompertz, Logistic and some their generalizations. The advantage is to obtain a single mathematically tractable equation from which the main characteristics of the considered curves can be deduced. We focus on the effects of the involved parameters through both analytical results and computational evaluations.
	\end{abstract}
\end{frontmatter}
%
\section{Introduction}
The need to describe and explain the evolution of many phenomena associated with growth curves has led to multiple efforts to study these types of functions. This need is more accentuated if we take into account the application of these curves in a wide variety of fields of application in branches of the physical, biological and social sciences, as well as in various areas of agriculture, business, education, engineering, medicine and public health (see, for example, Banks \cite{Ban94} where much of the traditional growth curves are presented together with real examples of application).

Within the family of growth curves, those of the sigmoidal type deserve a special section. The first to appear in the literature is the logistic curve, introduced by Verhulst in the context of Demography and the study of population dynamics, although it took until the second decade of the 20th century for it to be retaken. Today it is one of the most widely used models in areas such as innovation diffusion modeling (Giovannis and Skiadas \cite{Gio07}) or the exploitation of energy resources (Giovannis and Skiadas \cite{Gio99}). Together with the logistic curve, the Gompertz one is perhaps the most widely discussed in the literature on modeling growth phenomena. Introduced by Benjamin Gompertz to model the law of human mortality, currently the study of it has become very important, especially since its usefulness was demonstrated in the description of tumor growths (see for example, Ferrante et al. \cite{Fer00}), which has given rise to a great boom in the study of it.

However, these curves cannot always adequately represent various real sigmoidal behavior patterns, mainly due to their stiffness in some of their characteristics such as the inflection points. This led to the appearance of new curves, among which we highlight the one introduced by Von Bertalanffy or the later extension due to Richards (see Rom\'an-Rom\'an et al. \cite{Rom10} and Rom\'an-Rom\'an and Torres-Ruiz \cite{Rom15}, and references therein). Later generalizations gave rise to more general expressions such as the Hyperlogistic and Blumberg curves.

The search for greater flexibility in the curves that allows dealing with complex real situations has led many authors to develop a generalization methodology starting from a simple equation to understand the growth mechanism of a specific phenomenon. Then, and with the aim of generating more flexible forms, and thereby increasing the applications to a wider range of research areas, more parameters, or functions, are incorporated into the model. One of the precursors of this type of study was Turner et al. \cite{Tur76} who propose a general theory of growth based on quite general postulates. Subsequent analyzes in this regard have been developed by Tsoularis and Wallace \cite{Tso02} and by Koya and Goshu \cite{Koy13}. Although generalized equations offer greater flexibility, the complexity of the model increases. For this reason, within this line, in recent years the idea of adding new parameters has been combined with that of introducing functions with very flexible behaviors. Tabatabai et al. \cite{Tab05} construct the so-called hyperbolastic curves that allow the adjustment of data showing different types of sigmoidal behavior, applying them to the study of tumor evolution and stem cell growth (Tabatabai et al. \cite{Tab11}). These models are the starting point for the development of others, such as the oscillabolastic one, which aim is to model oscillatory growths (Eby and Tabatabai \cite{Eby14}) and the T-type model (Tabatabai et al. \cite{Tab13}), capable of represent biphasic sigmoidal growths. Following this line, Erto and Lepore \cite{Ert20} have introduced a new type of sigmoidal curve that can, under certain conditions, present more than one inflection point. With regard to multisigmoidal curves, recently the works of Di Crescenzo et al. \cite{Cre21} and Rom\'an-Rom\'an et al. \cite{Rom19} have introduced multisigmoidal logistic and Gompertz models modifying the original ordinary differential equations by introducing polynomial functions in their expression. This is one of the procedures followed by various authors for the generalization of certain growth curves. For example, the hyperbolastic model H1 arises from the modification of the logistic equation, while the curve H3 appears as the solution of a modification of the Weibull differential equation. In both cases, the introduced  functions are of the hyperbolic type. Another way of introducing generalizations is from the alteration of stochastic models associated with the curves. Along these lines, we can highlight the introduction of temporal functions to model the incorporation of therapies in the evolution of tumor growths, giving rise to modifications of the Gompertz curve (see for example Albano et al. \cite{Alb11,Alb13,Alb15,Alb20}).

A fundamental aspect to take into account is that all these modifications are due to the need to have the broadest possible knowledge of the curves considered. An example of great relevance today is the study of the evolution of epidemics such as Covid19, where it is of fundamental interest to determine the instants of contagion peaks, the moments of inflection of the evolution, duration times and of the appearance of successive waves. In this sense, the objective of this article is to analyze in depth a general deterministic model of growth that includes and generalizes the most widely used models. Specifically, we consider the generazed form of the logistic curve introduced in Tsoularis and Wallace \cite{Tso02}. Such growth form incorporates also growth models different from the logistic growth and its generalizations, making it able to fit population dynamics in a very wide range of behaviors. This is essentially the reason why in this paper we investigate the role of some relevant parameters in the equation governing the growth in Tsoularis and Wallace \cite{Tso02} and we analitycaly investigate the behavior of the curve in different scenarios.\par
The plan of the paper is the following. In Section~\ref{Section_2}, starting from Tsoularis and Wallace \cite{Tso02},  a general ordinary differential equation  to describe the evolution of a population size is considered and the solution  is determined. We also show that  choosing appropriately the parameters one can derive the more used models; specifically we show that the Malthus, Hyper-logistic, logistic, Hyper-Gompertz, Gompertz, Bertalanffy-Richards models can be obtained from the general curve.
In Section~\ref{Section_3} an analytical study of the considered curve is performed and we analyze the various behaviors that the curve can exhibit. In Section~\ref{Section_4} we provide a detailed numeric analysis apt to show the flexibility of the model based on the choice of parameters. Some concluding remarks close the paper.

\section{The model}\label {Section_2}
We denote by $x=x(t)$ the population size or the dimension of an organism at time $t$. We assume that $x(t)$ is the solution of the following ordinary differential equation (ODE)
\begin{equation}\label{ODE}
\dfrac{dx}{dt}=\gamma k^{n(p-1)}x^{1+n(1-p)}\left[1-\left(\dfrac{x}{k}\right)^n\right]^{p}, \quad x(t_0)=x_0,
\end{equation}
where $x_0$ denotes the population size at the initial time $t_0$, $\gamma$, $n$ and $p$ are shape-parameters subject to being  positive with $0<p<1+1/n$ and $k>0$. As we will see later, this last parameter, usually interpreted as the carrying capacity of the system in the classical sigmoidal curves, here has different interpretations depending on the other parameteres.
\begin{remark}\label{remark1}
The most used ODE's considered as growth equations can be expressed in the form given in \eqref{ODE}. In particular, the following cases are included in \eqref{ODE}.
\begin{itemize}
\item {\bf Bertalanffy-Richards:} For $p\rightarrow 1$, Eq. \eqref{ODE} reduces to
\begin{equation}\label{ODE_ Richards}
\dfrac{dx}{dt}=\gamma x\left[1-\left(\dfrac{x}{k}\right)^n\right], \quad x(t_0)=x_0.
\end{equation}
\item {\bf Logistic:}  Eq. \eqref{ODE_ Richards} becomes the classical logistic equation when $n\rightarrow 1$:
\begin{equation}\label{ODE_Logistic}
\dfrac{dx}{dt}=\gamma x\left(1-\dfrac{x}{k}\right), \quad x(t_0)=x_0.
\end{equation}
\item {\bf Malthus:}  For $k\rightarrow \infty$, Eq. \eqref{ODE_ Richards} leads to Malthus equation:
$$
\dfrac{dx}{dt}=\gamma x, \quad x(t_0)=x_0.
$$
\item {\bf Hyper-Logistic:}  When $n\rightarrow 1$, from \eqref{ODE} we obtain:
$$
\dfrac{dx}{dt}=\gamma k^{p-1}x^{2-p}\left(1-\dfrac{x}{k}\right)^{p}, \quad x(t_0)=x_0,
$$
that includes the Logistic ODE \eqref{ODE_Logistic} obtainable when $p\rightarrow 1$.
\item {\bf Hyper-Gompertz}: From Eq. \eqref{ODE} we derive the Hyper-Gompertz equation for $n\rightarrow 0$. Indeed, since
 $\lim\limits_{n \to 0}\dfrac{k^n-x^n}{n}=\ln\left(\dfrac{k}{x}\right)$, assuming that $\lim\limits_{n \to 0}\gamma n^p=\gamma'$, from \eqref{ODE} we obtain
 \begin{equation}\label{ODE_Hypergompertz}
\dfrac{dx}{dt}=\gamma' x\left[\ln\left(\dfrac{k}{x}\right)\right]^{p}.
\end{equation}
\item {\bf Gompertz:} Considering the limit  $p\rightarrow 1$, Eq.~\eqref{ODE_Hypergompertz}   becomes the simplest Gompertz growth curve:
$$
\dfrac{dx}{dt}=\gamma' x\ln\left(\dfrac{k}{x}\right).
$$
\end{itemize}
\end{remark}
Discussion of Remark~\ref{remark1} is summarized in the scheme in Figure \ref{Schema}.
\tikzstyle{startstop} = [ellipse, rounded corners, minimum width=3cm,
minimum height=1cm,text centered, draw=black, fill=red!30]
\tikzstyle{io} = [trapezium, trapezium left angle=70, trapezium right angle=110, minimum width=3cm,
minimum height=1cm, text centered, draw=black, fill=blue!30]
\tikzstyle{process} = [rectangle, minimum width=1.5cm,
minimum height=1cm, text centered, draw=black, fill=white!30]
\tikzstyle{decision} = [diamond, minimum width=1.5cm,
minimum height=1cm, text centered, draw=black, fill=green!30]
\tikzstyle{arrow} = [thick,->,>=stealth]
\tikzstyle{unione} = [circle, rounded corners, minimum width=0.1cm,
minimum height=0.1cm,text centered, draw=black, fill=red!30]
\begin{figure}[!ht]
  \centering
\begin{tikzpicture}[node distance=2cm]
 \node[text width=3cm] at (10,0) {};
\node (GC)  [process] {
\begin{minipage}{0.45\textwidth}
\footnotesize$\begin{array}{c}{\rm General \;growth\; equation}\\
\footnotesize \frac{dx(t)}{dt}=\gamma\,k^{n(p-1)}x^{1+n(1-p)}\left[1-\left(\frac{x}{k}\right)^n\right]^p\end{array}$
\end{minipage}
};
\node[text width=3cm,yshift=-4.5cm] at (10,0)  {};
\node (BR) [process, below of=GC,yshift=-1cm] {
\begin{minipage}{0.38\textwidth}
\footnotesize$\begin{array}{c}{\rm  Bertalanffy-Richards\; equation}\\
\footnotesize\frac{dx(t)}{dt}=\gamma x\left[1-\left(\frac{x}{k}\right)^n\right]\end{array}$
\end{minipage}
};
\node[text width=3cm,yshift=-4cm] at (10,0)  {};
\node (HG) [process,left of=BR,xshift=-3.3cm] {
\begin{minipage}{0.34\textwidth}
\footnotesize$\begin{array}{c} {\rm Hyper-Gompertz\; equation}\\
\footnotesize \frac{dx(t)}{dt}=\gamma'x\left[\ln\left(\frac{k}{x}\right)\right]^p\end{array}$\end{minipage}
};
\node[text width=3cm,yshift=-4cm] at (10,0)  {};
\node (HL) [process, right of=BR,xshift=3.3cm] {
\begin{minipage}{0.32\textwidth}
\footnotesize$\begin{array}{c}{\rm Hyper-logistic\; equation}\\
\footnotesize \frac{dx(t)}{dt}=\gamma k^{p-1}x^{2-p}\left(1-\frac{x}{k}\right)^p\end{array}$
\end{minipage}
};
\node[text width=3cm,yshift=-4cm] at (10,0)  {};
\node (G) [process,below of=HG,yshift=-1cm] {
\begin{minipage}{0.23\textwidth}
\footnotesize$\begin{array}{c} {\rm Gompertz\; equation}\\
\footnotesize  \frac{dx(t)}{dt}=\gamma'x\ln\left(\frac{k}{x}\right)\end{array}$
\end{minipage}
};
\node[text width=3cm,yshift=-4cm] at (10,0)  {};
\node (L) [process, below of=HL,yshift=-1cm] {
\begin{minipage}{0.22\textwidth}
\footnotesize$\begin{array}{c} {\rm Logistic\; equation}\\
\footnotesize\frac{dx(t)}{dt}=\gamma x\left(1-\frac{x}{k}\right)\end{array}$
\end{minipage}
};
\node[text width=3cm,yshift=-4cm] at (10,0)  {};
\node (M) [process,below of=BR,yshift=-3cm] {
\begin{minipage}{0.2\textwidth}
\footnotesize$\begin{array}{c} {\rm Malthus\; equation}\\
\footnotesize \frac{dx(t)}{dt}=\gamma x\end{array}$
\end{minipage}
};
\path[->] (GC) edge node {\hspace{-2.5cm}\scriptsize$\begin{array}{c}n\to 0\\\gamma n^p\to\gamma'\end{array}$} (HG);
\path[->] (GC) edge node {\hspace{1.cm}\scriptsize $p\to 1$} (BR);
\path[->](GC) edge node {\hspace{1.5cm}\scriptsize $n=1$} (HL);
\path[->] (BR) edge node {\hspace{1.8cm}\scriptsize $\begin{array}{c}n\to 0\\\gamma n^p\to\gamma'\end{array}$} (G);
\path[->] (BR) edge node {\hspace{1.5cm}\scriptsize $n\to 1$} (L);
\path[->](HG) edge node {\hspace{1.cm}\scriptsize $p\to 1$} (G);
\path[->] (HL) edge node {\hspace{1.cm}\scriptsize $p\to 1$} (L);
\path[->] (BR) edge node {\hspace{1.2cm}\scriptsize $k\to \infty$} (M);
\path [->](G) edge node  {\hspace{-2.5cm}\scriptsize $\begin{array}{c}k\to \infty\\\gamma'\ln k=\gamma\end{array}$} (M);
\path[->](L) edge node {\hspace{2.cm}\scriptsize $k\to \infty$}(M);
\end{tikzpicture}
\caption{Scheme summarizing the results of Remark \ref{remark1}.}
\label{Schema}
\end{figure}
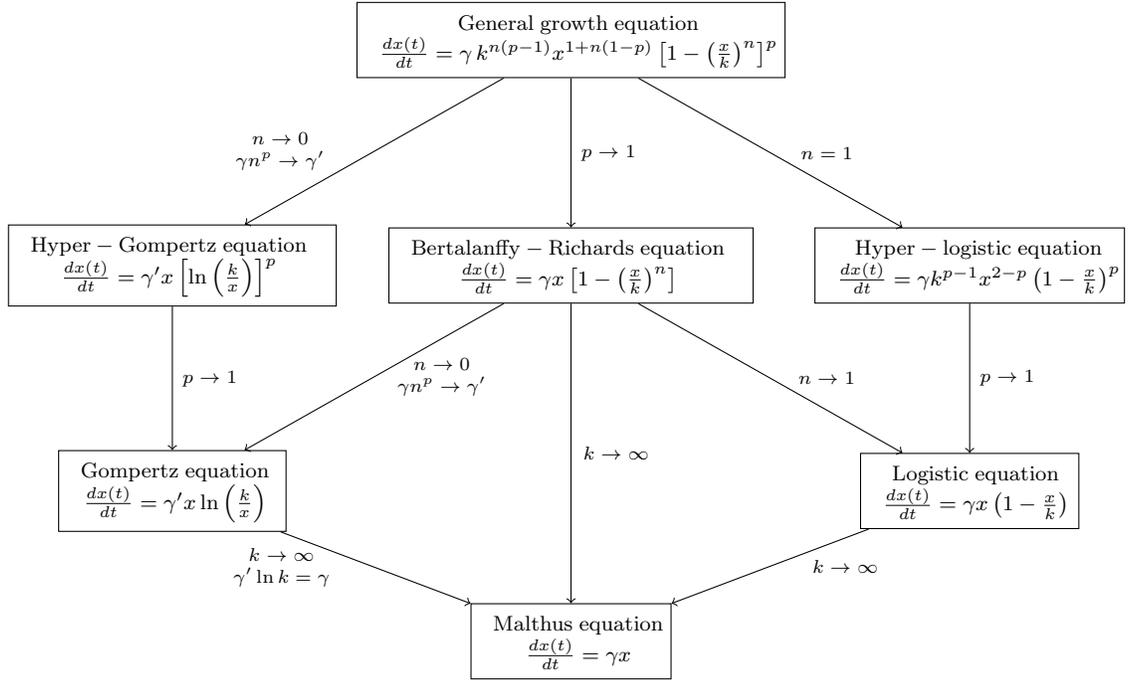
\begin{lemma} The solution of Eq. \eqref{ODE} is
\begin{equation}\label{sol1}
x(t)=\dfrac{k}{\left\{1+\left[\gamma n(p-1)(t-t_0)+A_n^{1-p}\right]^{\frac{1}{1-p}}\right\}^{1/n}},
\end{equation}
where $A_n=\left(\dfrac{k}{x_0}\right)^{n}-1$ depends on the shape parameter $n$ and on the ratio between the carrying capacity $k$ and the initial population size $x_0$.
\end{lemma}
\begin{proof}
Integrating both the members in \eqref{ODE} we obtain:
\begin{equation}\label{int1}
\int^t x^{-1-n(1-p)}\left[1-\left(\dfrac{x}{k}\right)^n\right]^{-p} dx = \gamma k^{n(p-1)} t + C,
\end{equation}
where $C$ is a constant of integration.
We consider the integral on the left hand size of \eqref{int1}. Carrying out the change of the variable of integration $x$ in
$$v=\left(\dfrac{k}{x}\right)^n-1,$$
so that
$$dx= -\frac{k}{n}\, \frac{1}{(v+1)^{-(1+n)/n}},$$
we obtain the solution of the integral in \eqref{int1}:
\begin{equation}
\label{int2}
\int x^{-1-n(1-p)}\left[1-\left(\dfrac{x}{k}\right)^n\right]^{-p} dx =-\frac{k^{-n(1-p)}}{n} \int v^{-p} dv =\dfrac{k^{-n(1-p)}}{n(p-1)}v^{-p+1}.
\end{equation}
From \eqref{int2}, by making use of the inverse transformation, it follows:
\begin{equation}
\label{int3}
\int x^{-1-n(1-p)}\left[1-\left(\dfrac{x}{k}\right)^n\right]^{-p} dx =\frac{k^{-n(1-p)}}{n(p-1)}\left(\dfrac{k^n-x^n}{x^n}\right)^{1-p}.
\end{equation}
The constant $C$ in Eq. \eqref{int1} can be obtained by using the initial condition $x(t_0)=x_0$ resulting in
\begin{equation}
\label{C}
C=\dfrac{k^{n(p-1)}}{n(p-1)}\left(\dfrac{k^n-x_0^n}{x_0^n}\right)^{1-p}-\gamma k^{n(p-1)} t_0.
\end{equation}
Therefore, making use of \eqref{int3} and \eqref{C} in \eqref{int1}, we obtain
\begin{equation*}
\dfrac{k^{n(p-1)}}{n(p-1)}\left[\left(\dfrac{k^n-x^n}{x^n}\right)^{1-p}-\left(\dfrac{k^n-x_0^n}{x_0^n}\right)^{1-p}\right]=\gamma k^{n(p-1)}(t-t_0).
\end{equation*}
or, equivalently,
\begin{equation*}
\left(\dfrac{k}{x}\right)^n=1+\left[\gamma n(p-1)(t-t_0)+A_n^{1-p}\right]^{\frac{1}{1-p}},
\end{equation*}
from which \eqref{sol1} follows.
$\hfill\square$
\end{proof}
On the line with the Remark {\ref{remark1}}, in the following remark we show that the solution in \eqref{sol1} includes the most used growth curves.
\begin{remark} \label{Remark2}
The equation \eqref{sol1} includes the most used equations in literature. In particular, the following cases are included.
\begin{itemize}
\item {\bf Bertalanffy-Richards:}
When $p\rightarrow 1$, Eq. \eqref{sol1} reduces to Bertalanffy-Richards curve. Indeed, since
$$\lim_{p\to 1} \left\{1+\left[\gamma n(p-1)(t-t_0)+A_n^{1-p}\right]^{\frac{1}{1-p}}\right\}=A_n e^{-\gamma n (t-t_0)},$$
from \eqref{sol1} one has:
\begin{equation}\label{sol_ Richards}
x(t)=\frac{k}{\bigl[1+A_ne^{-\gamma n (t-t_0)}\bigr]^{1/n}}.
\end{equation}
\item {\bf Logistic:}  For $n\rightarrow 1$, Eq. \eqref{sol_ Richards} becomes the classical logistic curve:
\begin{equation}\label{sol_Logistic}
x(t)=\frac{k}{\bigl[1+A_1e^{-\gamma (t-t_0)}\bigr]}.
\end{equation}
\item {\bf Malthus:}  We point out that, for $k\rightarrow \infty$, Eq. \eqref{sol_Logistic} reduces to Malthus curve, i.e.
$$
x(t)=x_0 e^{\gamma (t-t_0)}.
$$
\item {\bf Hyper-Logistic:}  When $n\rightarrow 1$, from \eqref{sol1} we obtain the hyper-logistic curve:
$$
x(t)=\dfrac{k}{1+\left[\gamma (p-1)(t-t_0)+A_1^{1-p}\right]^{\frac{1}{1-p}}}.
$$
that for $p\rightarrow 1$ becomes the logistic growth.
\item {\bf Hyper-Gompertz:}   From \eqref{sol1} we derive the Hyper-Gompertz equation. Indeed, since
\begin{eqnarray}
&&\hspace{-1cm}\lim_{n \to 0} \frac{\left[\gamma n(p-1)(t-t_0)+A_n^{1-p}\right]^{\frac{1}{1-p}}}{n}\nonumber\\
&&\hspace{0.8cm}=\lim_{n \to 0} \left[\frac{\gamma n(p-1)(t-t_0)+A_n^{1-p}}{n^{1-p}}\right]^{\frac{1}{1-p}}\nonumber\\
&&\hspace{0.8cm}=\lim_{n \to 0}\left[\gamma n^p(p-1)(t-t_0)+\left(\frac{k^n-x_0^n}{nx_0^n}\right)^{1-p}\right]^{\frac{1}{1-p}},\nonumber
\end{eqnarray}
so, assuming $\gamma n^p \longrightarrow \gamma'$, we have:
\begin{eqnarray}
&&\hspace{-1cm}\lim_{n \to 0}\left[\gamma n^p(p-1)(t-t_0)+\left(\dfrac{k^n-x_0^n}{nx_0^n}\right)^{1-p}\right]^{\frac{1}{1-p}}\nonumber\\
&&=\left[\gamma'(p-1)(t-t_0)+\left[\ln\left(\dfrac{k}{x_0}\right)\right]^{1-p}\right]^{\frac{1}{1-p}}.\nonumber
\end{eqnarray}
Therefore, we conclude that
\begin{equation}
\label{HyperGompertz}
\lim\limits_{n \to 0}x(t)=k\exp\left(-\left[\gamma'(p-1)(t-t_0)+\left(\ln\left(\dfrac{k}{x_0}\right)\right)^{1-p}\right]^{\frac{1}{1-p}}\right)
\end{equation}
corresponding to the Hyper-Gompertz curve.
\item {\bf Gompertz:}  Taking the limit for $p\rightarrow 1$ in Eq.~\eqref{HyperGompertz} we obtain the  Gompertz curve. Indeed, since
\begin{equation*}
\lim_{p \to 1}\left[\gamma'(p-1)(t-t_0)+\left(\ln\left(\dfrac{k}{x_0}\right)\right)^{1-p}\right]^{\frac{1}{1-p}}=\ln\left(\dfrac{k}{x_0}\right)e^{-\gamma(t-t_0)},
\end{equation*}
we conclude
\begin{equation*}
\lim_{\substack{n \to 1\\p \to 1}}x(t)=k\exp\left(-\ln\left(\dfrac{k}{x_0}\right)e^{-\gamma(t-t_0)}\right),
\end{equation*}
that is the Gompertz curve.
\end{itemize}
\end{remark}
In the following section we will study the growth curve in \eqref{sol1}, analyzing the domain, the monotonicity and the potential inflection points and the role of the parameters on such issues. In particular, we consider the parameters $\gamma, x_0$ and $k$ as fixed and we explore the role of $n$ and $p$ in \eqref{sol1}.
%
%
\section{Analysis of the growth curve}\label{Section_3}
We rewrite Eq. \eqref{sol1} in the following general form:
\begin{equation}\label{sol2}
x(t)=\dfrac{k}{\left[1+g_\theta(t)^{\frac{1}{1-p}}\right]^{1/n}}.
\end{equation}
where
\begin{equation}\label{def_g}
g_\theta(t):=\gamma n(p-1)(t-t_0)+A_n^{1-p}.
\end{equation}
%
It is interesting to evaluate the time point $t_S$ in which the population size reaches a fixed value $S<k$. Such value can be a proportion of the the carrying capacity $k$ or a multiplier of the initial size $x_0$.
In particular, from \eqref{sol2} and \eqref{def_g}, it is easy to see that
$$
t_S=t_0+\frac{\bigl[\bigl(\frac{k}{S}\bigr)^n-1\bigr]^{1-p}-A_n^{1-p}}{\gamma n (p-1)}.
$$
To analyze the monotonicity of the function $x(t)$, we evaluate the first derivative of Eq. \eqref{sol2}:
\begin{equation}\label{Derivata_x}
\dfrac{dx}{dt}=x(t)h_\theta(t),
\end{equation}
where
\begin{equation}\label{Def_h}
h_\theta(t)=\gamma \frac{\left[g_\theta(t)\right]^\frac{p}{1-p}}{1+\left[g_\theta(t)\right]^\frac{1}{1-p}}.
\end{equation}
In order to determine the potential inflection points of $x(t)$, we consider the second derivative:
\begin{equation}\label{Derivata2_x}
\dfrac{d^2x}{dt^2}=\dfrac{dx}{dt}h_{\theta}(t)+x(t)\dfrac{dh_{\theta}(t)}{dt}=
x(t)\left[\dfrac{dh_{\theta}(t)}{dt}+\left[h_{\theta}(t)\right]^2\right].
\end{equation}
From \eqref{Def_h}, we have
$$
\dfrac{dh_{\theta}(t)}{dt}=\dfrac{n\,\gamma^2 [g_{\theta}(t)]^\frac{2p-1}{1-p}}{\left(1+[g_{\theta}(t)]^\frac{1}{1-p}\right)^2}
\left\{[g_{\theta}(t)]^\frac{1}{1-p}-p\left(1+[g_{\theta}(t)]^\frac{1}{1-p}\right)\right\},
$$
so that
\begin{eqnarray}\label{nonso}
&& \hspace{-1.0cm}\frac{dh_{\theta}(t)}{dt}+\left[h_{\theta}(t)\right]^2\nonumber\\
&& =\frac{\gamma^2 [g_{\theta}(t)]^\frac{2p-1}{1-p}}{\left(1+[g_{\theta}(t)]^\frac{1}{1-p}\right)^2}\left\{n [g_{\theta}(t)]^\frac{1}{1-p}-np\left(1+[g_{\theta}(t)]^\frac{1}{1-p}\right)+[g_{\theta}(t)]^\frac{1}{1-p}\right\}\nonumber\\
&&=\frac{\gamma^2 [g_{\theta}(t)]^\frac{2p-1}{1-p}}{\left(1+[g_{\theta}(t)]^\frac{1}{1-p}\right)^2} \left\{[g_{\theta}(t)]^\frac{1}{1-p}\left[n(1-p)+1\right]-np\right\}.
\end{eqnarray}
In our analysis, we focus also on the proportion of the population in the point in which the growth velocity is maximum and its carrying capacity. Such proportion mathematically is the ratio $\pi_x:=\frac{x(t_{inf})}{k}$, where $t_{inf}$ represents the inflection point of the curve $x(t)$. We will see that $\pi_x$ depends only on the parameters $n$ and $p$.\par\noindent
%
%
In the following we distinguish three cases:
\begin{enumerate}
\item $1<p<1+\frac{1}{n}$;
\item $p=1$;
\item $0<p<1$.
\end{enumerate}
%
\subsection{Analysis of the curve in the Case 1: $1<p<1+\frac{1}{n}$}
Since in such case  $g_\theta(t)>0$ for all $t\in [t_0, \infty)$, the function $x(t)$ is defined in $[t_0,\infty)$. Further, being $\lim_{t\to +\infty}g_{\theta}(t)=+\infty$, the denominator of Eq. \eqref{sol2} tends to $1$, so $\lim_{t\to +\infty} x(t)=k$.
\noindent
To study the monotonicity of the function $x(t)$, we analyze the first derivative in \eqref{Derivata_x}. In particular, from \eqref{Def_h}, due to the sign of the function $g_{\theta}(t)$, we have that $x(t)$ is monotonically increasing.\\
\noindent
The inflection points are obtained by setting  $\dfrac{d^2x}{dt^2} =0$ in \eqref{Derivata2_x}.  Since $1<p<1+1/n$, we have $n(1-p)+1>0$, hence from \eqref{nonso}, an inflection point $ t_{Inf}$ exists if and only if
$$
g_{\theta}(t_{Inf})=\left[\frac{np}{1+n(1-p)}\right]^{1-p},
$$
from  which, recalling \eqref{def_g}, we obtain the following inflection point:
\begin{equation}
 t_{Inf}=t_0+\dfrac{\left[\dfrac{np}{1+n(1-p)}\right]^{1-p}-A_n^{1-p}}{n\gamma(p-1)}.
\label{flesso_caso1}
\end{equation}
We note that, since $p>1$, the inflection point is visible, i.e. $t_{Inf}\geq t_0$, if and only if
$$
\dfrac{np}{1+n(1-p)}\geq A_n
$$
i.e.
$$
p\leq \left(1+\frac{1}{n}\right)\frac{A_n}{A_n+1}= \left(1+\frac{1}{n}\right)\,\left[1-\left(\frac{x_0}{k}\right)^n\right].
$$
Moreover, from \eqref{ODE}   and \eqref{flesso_caso1}, we obtain:
$$
x( t_{Inf})=\frac{k}{\left[1+\frac{np}{1+n\,(1-p)}\right]^{\frac{1}{n}}}=k \left[\frac{1+n\,(1-p)}{1+n}\right]^{\frac{1}{n}}=k \left[1-\frac{np}{1+n}\right]^{\frac{1}{n}}.
$$
We note that the proportion between $x( t_{Inf})$ and the carrying capacity $k$ is
$$
\pi_x=\left[1-\frac{np}{1+n}\right]^{\frac{1}{n}}=\left[1+\frac{np}{n(1-p)+1}\right]^{-\frac{1}{n}}.
$$
%
%
\subsection{Analysis of the curve in the Case 2: $p=1$}
Since
$$
\lim_{p\to 1} g_{\theta}(t)^\frac{1}{1-p}=A_n e^{-\gamma n(t-t_0)},
$$
taking the limit of $x(t)$ for $p\to 1$ in the Eq. \eqref{sol1}, we obtain the well known Richards equation:
$$
x(t)=\dfrac{k}{\left\{1+A_n e^{-\gamma n(t-t_0)}\right\}^{1/n}}.
$$
In such case $x(t)$ is defined in $[t_0,\infty)$; the carrying capacity is $\lim_{t\to \infty} x(t)=k$. The function $x(t)$ is stricty monotonically increasing, as it is evident from \eqref{Derivata_x} and \eqref{Def_h}. Further, from \eqref{Derivata2_x} and \eqref{nonso},
$$
\lim_{p\to 1} \left\{\dfrac{dh_{\theta}(t)}{dt}+\left[h_{\theta}(t)\right]^2\right\}=\frac{A_n e^{\gamma n t_0}\gamma^2\left[A_n e^{\gamma n t_0}-e^{\gamma n t} n\right]}{\left(e^{\gamma n t}-A_n e^{\gamma n t_0}\right)^2}
$$
from which we derive that an inflection point exists if and only if $A_n e^{\gamma n t_0}-e^{\gamma n t} n=0$, i.e.
\begin{equation}\label{flesso_caso2}
t_{Inf}=t_0+\frac{1}{\gamma n}\log\left(\frac{A_n}{n}\right).
\end{equation}
From \eqref{flesso_caso2} it is clear that $t_{Inf}>t_0$ if and only if $A_n>n$, i.e.
$$
k>x_0 (n+1)^{\frac{1}{n}}.
$$
In such case, from \eqref{ODE}   and \eqref{flesso_caso2}, we obtain:
$$
x( t_{Inf})=k \left(\frac{1}{1+n}\right)^{\frac{1}{n}}.
$$
The proportion between $x( t_{Inf})$ and the carrying capacity $k$ is
$$
\pi_x=\left(1+n\right)^{-\frac{1}{n}}.
$$
\subsection{Analysis of the curve in the Case 3: $0<p<1$}
When $0<p<1$ we distinguish three different cases:
\begin{enumerate}
      \item[a.] $\dfrac{1}{1-p}=m\in\mathbb{N}$, with $m$ even.
      \item[b.] $\dfrac{1}{1-p}=m\in\mathbb{N}$, with $m$ odd.
	  \item[c.] $\dfrac{1}{1-p}=m\notin\mathbb{N}$.
\end{enumerate}
{\bf Case 3a: $\dfrac{1}{1-p}=m\in\mathbb{N}$}, with $m$ even. \\
\noindent In such case  the function $x(t)$ is defined in $[t_0,+\infty)$, being $\left[g_{\theta}(t)\right]^{m}\geq 0$ for all $t\geq t_0$. Further, since the denominator of Eq. \eqref{sol2} diverges as $t$ goes to $\infty$, one has $\lim_{t\to +\infty}x(t)=0$, hence the population is doomed to extinction. The sign of the first derivative \eqref{Derivata_x} depends on Eq. \eqref{Def_h}, that we rewrite as
$$
h_\theta(t)=\gamma \frac{\left[g_\theta(t)\right]^{m-1}}{1+\left[g_\theta(t)\right]^m}.
$$
Being $m-1$ odd, the sign of $h_\theta(t)$ is the same of the function $g_\theta(t)$; hence recalling Eq. \eqref{def_g}, one has that the $\frac{dx}{dt}>0$ if and only if
\begin{equation}\label{tStar}
t<t_*=t_0+\frac{m A_n^\frac{1}{m}}{\gamma n}.
\end{equation}
Therefore, in such case a maximum point presents coordinates $(t_*,k)$.\\
\noindent
Concerning the convexity, Eq. \eqref{nonso} becomes
\begin{equation}\label{der_2_mPari}
\dfrac{dh_{\theta}(t)}{dt}+\left[h_{\theta}(t)\right]^2=\dfrac{\gamma^2 [g_{\theta}(t)]^{m-2}}{\left(1+[g_{\theta}(t)]^m\right)^2} \frac{1}{m}\left\{(m+n)[g_{\theta}(t)]^m-n(m-1)\right\}
\end{equation}
so, the second derivative \eqref{Derivata2_x} vanishes if and only if either $g_{\theta}(t)=0$ or
$$
\left\{(m+n)[g_{\theta}(t)]^m-n(m-1)\right\}=0.
$$
Recalling \eqref{def_g} and that $g_{\theta}(t)$ vanishes in the maximum point of the function $x(t)$, we have that $\frac{d^2x}{dt^2}=0$
 if and only if $\left\{(m+n)[g_{\theta}(t)]^m-n(m-1)\right\}=0$. This last one admits two solutions being $m$ even. The two inflection points, $t_{Inf_1}$ and $t_{Inf_2}$ are:
$$
t_{Inf_1},t_{Inf_2}=t_0+m\frac{A_n^{\frac{1}{m}}\pm \left[\frac{n(m-1)}{n+m}\right]^\frac{1}{m}}{\gamma n}.
$$
We note that $t_{Inf_1}<t_{Inf_2}$, whereas the value of $x(t)$ in such inflection points is the same, concretely
$$
x(t_{Inf_1})=x(t_{Inf_2})=k\left[\frac{n+m}{m(n+1)}\right]^\frac{1}{n}.
$$
We also note that when
\begin{equation}\label{condizione_flesso}
k>x_0\left[\frac{m(n+1)}{n+m}\right]^\frac{1}{n}
\end{equation}
both the inflection points are in $[t_0,\infty)$, otherwise $t_{Inf_1}<t_0$.\\
\noindent
Finally, the proportion between $x( t_{Inf}):=x(t_{Inf_1})=x(t_{Inf_2})$ and the carrying capacity $k$ is
\begin{equation}\label{proporzioneCaso3}
\pi_x=\left[\frac{n+m}{m(n+1)}\right]^\frac{1}{n}.
\end{equation}
{\bf Case 3b: $\dfrac{1}{1-p}=m\in\mathbb{N}$}, with $m$ odd. \\
\noindent In such case  the denominator of \eqref{sol2} is equal to zero for
\begin{equation}\label{t1}
t_1=t_0+\frac{m\left(1+A_n^{\frac{1}{m}}\right)}{\gamma n},
\end{equation}
so the function $x(t)$ is defined in $[t_0,t_1)$ and a vertical asymptote is present in $t_1$. In particular $\lim_{t\to t_1} x(t)=+\infty$, hence the population explodes at the finite time $t_1$.
Further, looking at \eqref{Def_h}, being $m=\frac{1}{1-p}$ odd, $\left[g_\theta(t)\right]^{m-1}>0$. Still, the denominator of \eqref{sol2} is positive in the interval $[t_0,t_1)$. Hence the function $x(t)$ is monotonically increasing in $[t_0,t_1)$.
\\
\noindent
Finally, from Eq. \eqref{der_2_mPari}  we conclude that $\frac{d^2 x}{dt}=0$ admits two solutions; the first one obtained by setting $\left\{(m+n)[g_{\theta}(t)]^m-n(m-1)\right\}=0$ and the second one is the solution of $g_{\theta}(t)=0$, i.e.
\begin{equation}\label{der2_casoPari}
t_{Inf_1}=t_0+m\frac{A_n^{\frac{1}{m}}- \left[\frac{n(m-1)}{n+m}\right]^\frac{1}{m}}{\gamma n},\qquad t_{Inf_2}=t_0+\frac{mA_n^\frac{1}{m}}{\gamma n}, \quad t_{Inf_1}<t_{Inf_2}.
\end{equation}
We note that $t_0<t_{Inf_1}<t_{Inf_2}<t_1$ if and only if \eqref{condizione_flesso} holds.\par\noindent
Further,
\begin{equation}\label{Caso3b}
x(t_{Inf_1})=k\left[\frac{n+m}{m(n+1)}\right]^\frac{1}{n}, \qquad x(t_{Inf_2})=k
\end{equation}
and $\pi_x$ is given in \eqref{proporzioneCaso3}.\par
\noindent
%
%
%
{\bf Case 3c: $\dfrac{1}{1-p}=m\notin\mathbb{N}$}. \\
\noindent In this case  the function $x(t)$ is defined in $[t_0,t_2)$, with
\begin{equation}\label{t2}
t_2=t_0+\frac{m A_n^\frac{1}{m}}{\gamma n}.
\end{equation}
Further, $\lim_{t\to t_2} x(t)=k$, hence the population reaches its carrying capacity at the finite time $t_2$. From \eqref{Derivata_x} and \eqref{Def_h} we observe that $x(t)$ is monotonically increasing since $h(t)>0$ in $[t_0, t_2)$. Finally, since Eq. \eqref{nonso} vanishes in $t_{Inf}\equiv t_{Inf_1}$ as in \eqref{der2_casoPari}, we find an inflection point in
$$
t_{Inf}=t_0+m\frac{A_n^{\frac{1}{m}}- \left[\frac{n(m-1)}{n+m}\right]^\frac{1}{m}}{\gamma n}; \qquad x(t_{Inf})\equiv k\left[\frac{n+m}{m(n+1)}\right]^\frac{1}{n},
$$
while $\pi_x$ is given in \eqref{proporzioneCaso3}.
\section{Numerical analysis}\label{Section_4}
In this section we perform a numerical analysis of the curve \eqref{sol2}, by considering the cases identified in Section \ref{Section_3}. For each case, we analyze the behavior of the curve as function of the time, for different choices of the parameter $p$ by fixing $n$ and for several choices of $n$ by fixing $p$. The other parameters are fixed for all the study as: $t_0=0, x_0=1, \gamma=0.5$  and $k=400$.
\par
\begin{figure}[htbp]
\centering
\includegraphics[scale=0.4]{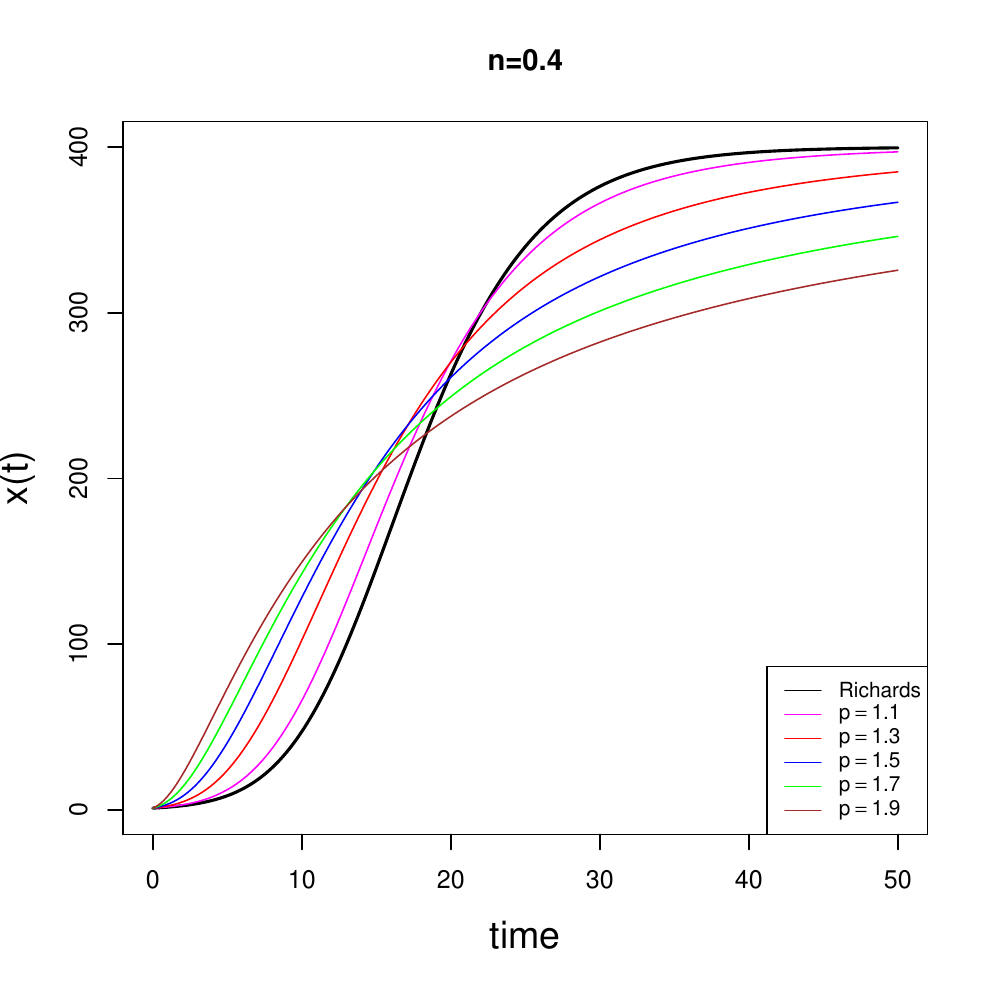}\includegraphics[scale=0.4]{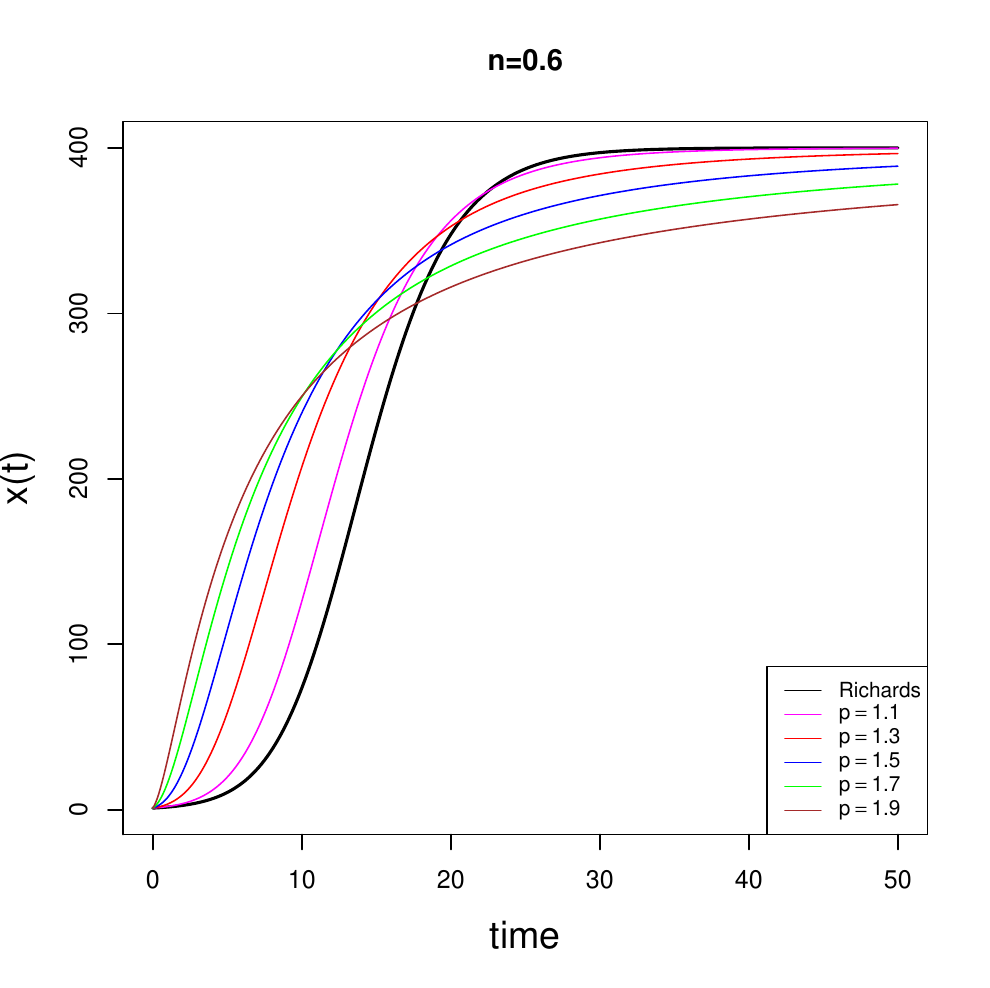}\\
\includegraphics[scale=0.4]{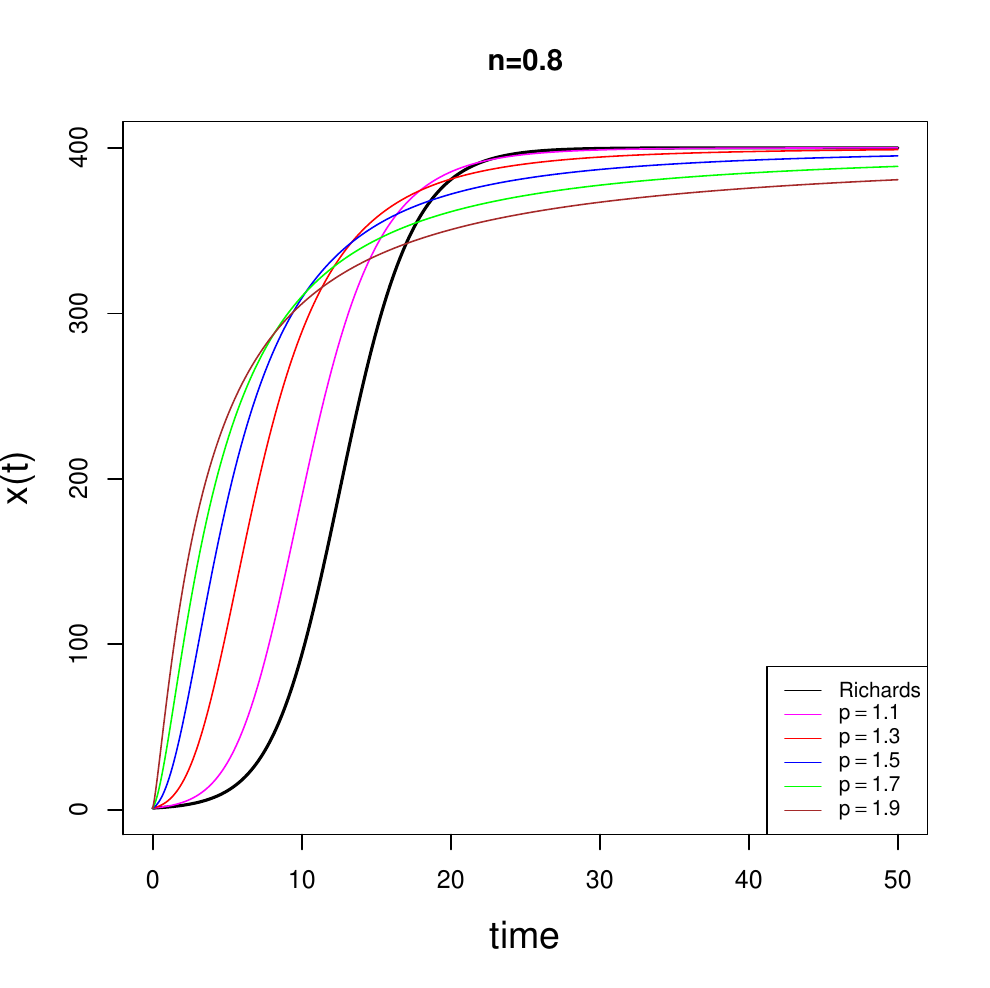}\includegraphics[scale=0.4]{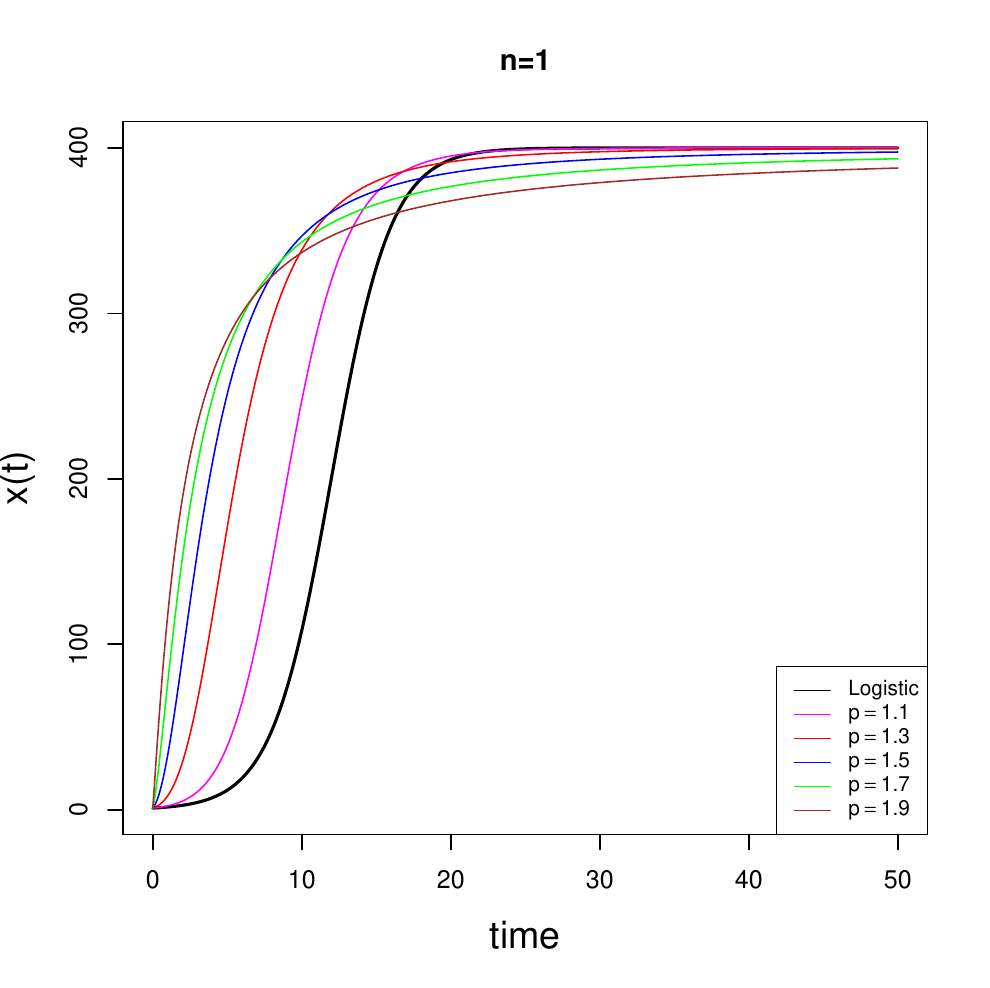}
\caption{Case 1: for several choices of the parameter $p$ with $1<p<1+\frac{1}{n}$, the general curve \eqref{sol2}  is  compared with the Richard curve (black line), fixed $n=0.4$, $n=0.6$, $n=0.8$ and $n=1$. In the last case the Richard curve becomes the logistic curve.}\label{fig2}
\end{figure}
\begin{figure}[htbp]
\centering
\includegraphics[scale=0.4]{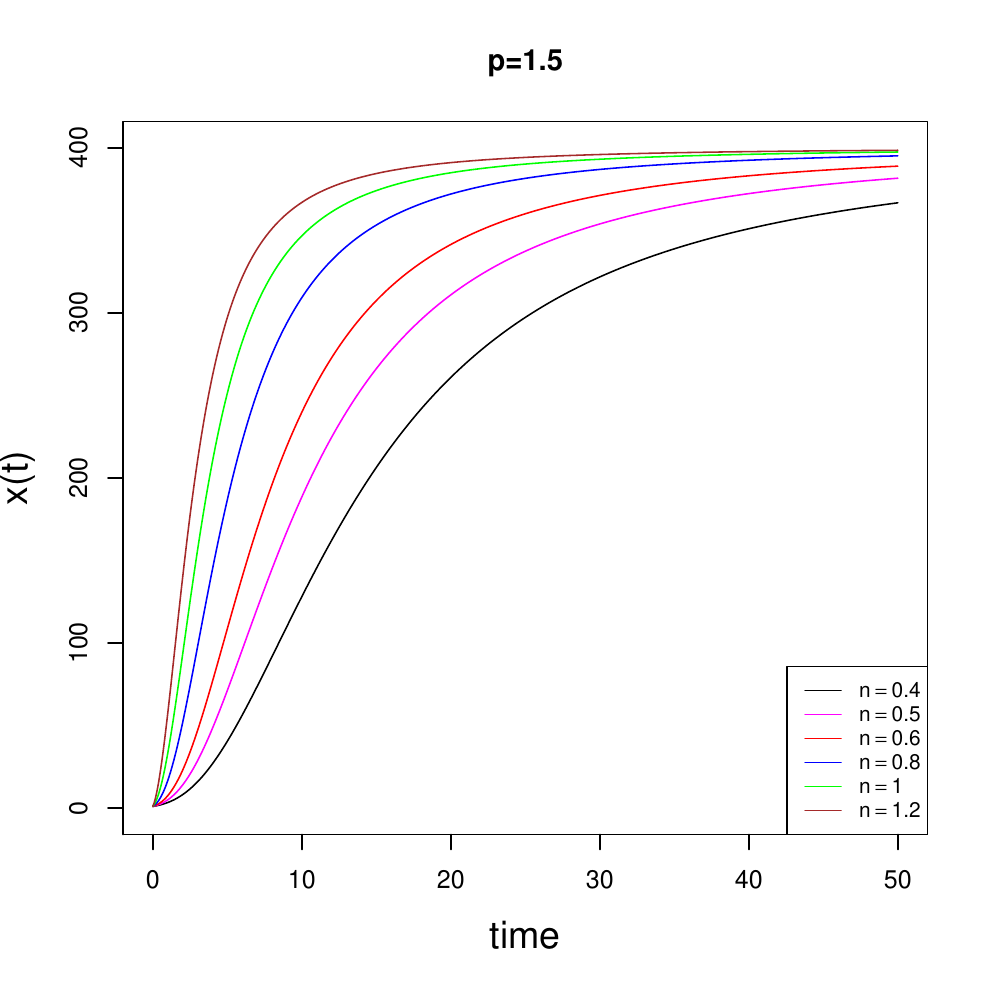}\includegraphics[scale=0.4]{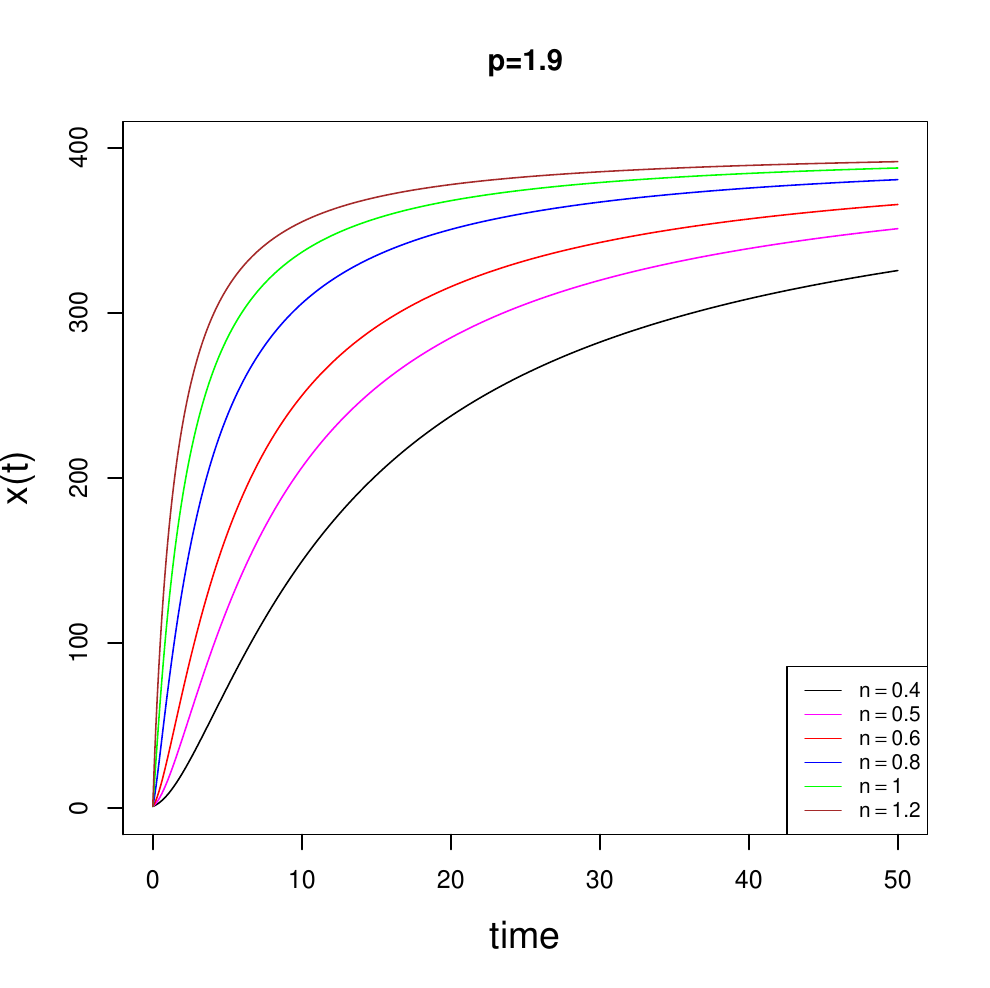}\\
\caption{Case 1: for  $p=1.5$ (on the left) and $p=1.9$ (on the right), the general curve \eqref{sol2} is  plotted for several choices of $n$ with $n<\frac{1}{1-p}$.}\label{fig3}
\end{figure}

In Figure \ref{fig2} the general curve in the case $1<p<1+\frac{1}{n}$ is compared with the Richards curve (black line), fixed $n=0.4, 0.6,0.8$ and $n=1$ and for several choices of the parameter $p$ ranging from $1.1$ to $1.9$. We point out that Richards curve is obtained from \eqref{sol2} when $p$ tends to 1. We can see that, by controlling $n$, the carrying capacity $k=400$ is reached in a time interval that is increasing as $p$ increases, remarking that the parameter $p$ has the role of proliferation rate for the population $x$.  Further, by increasing $n$ for each curve the carrying capacity is reached faster. This is also clear in Figure \ref{fig3} in which we plot the curve \eqref{sol2} in the Case 1 for several choice of $n$ ranging from $0.4$ to $1.2$ and for $p=1.5$ and $p=1.9$. From Figures \ref{fig2} and \ref{fig3} it is also clear that the inflection point $t_{Inf}$ calculated in \eqref{flesso_caso1} decreases as $p$ and $n$ increase. In order to give a quantitative evaluation of how the inflection point $t_{Inf}$ and $x(t_{Inf})$ change as $p$ and $n$ change, in Table \ref{TABELLA} such values are shown for $n=0.4, 0.6, 0.8$ and $1$ from left to right and for several choices of $p$ ranging from $1.1$ to $1.9$. Clearly, since the curve presents a unique inflection point, we set $t_{Inf_1}:=t_{Inf}$. The proportion $\pi_x$ between the value $x(t_{Inf_1})$ and the carrying capacity of curve is also listed in order to give a measure of the curve in which the growth velocity is maximum.\par
%
%
%
%
%
\begin{figure}[htbp]
\centering
\includegraphics[scale=0.35]{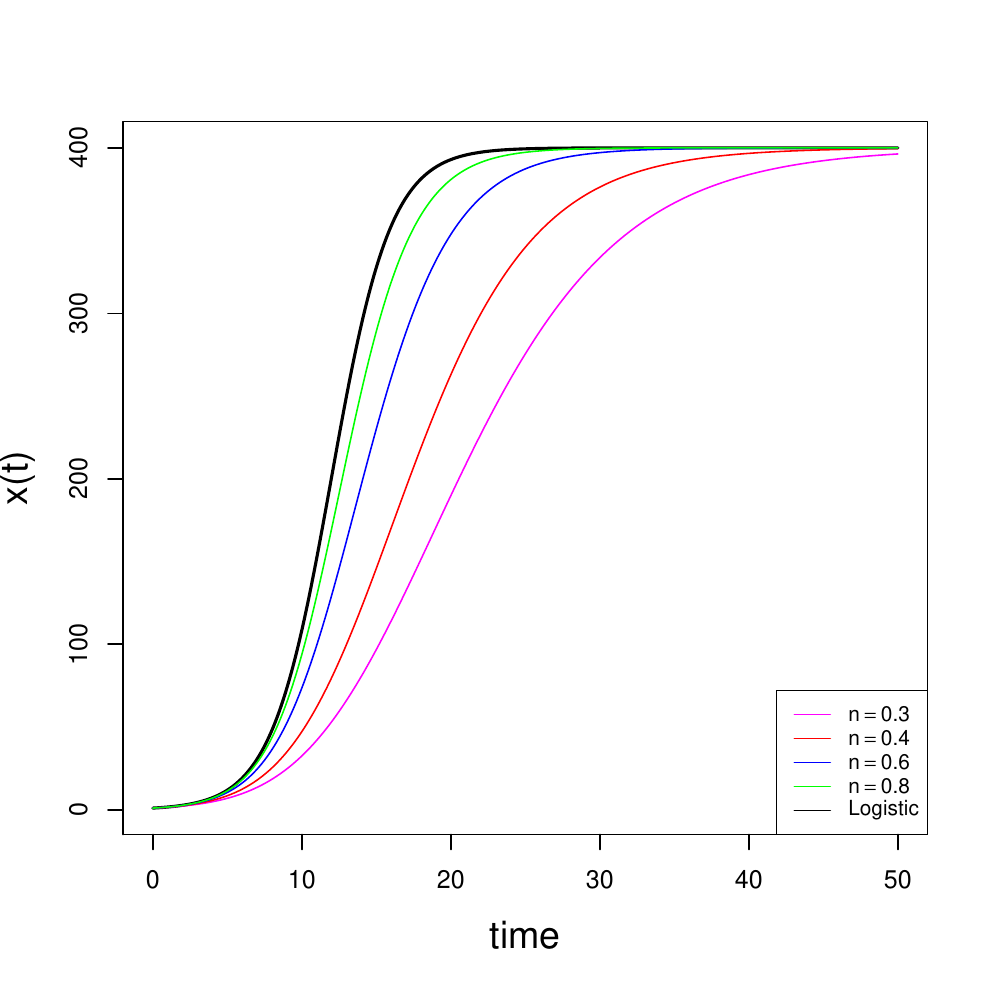}\includegraphics[scale=0.35]{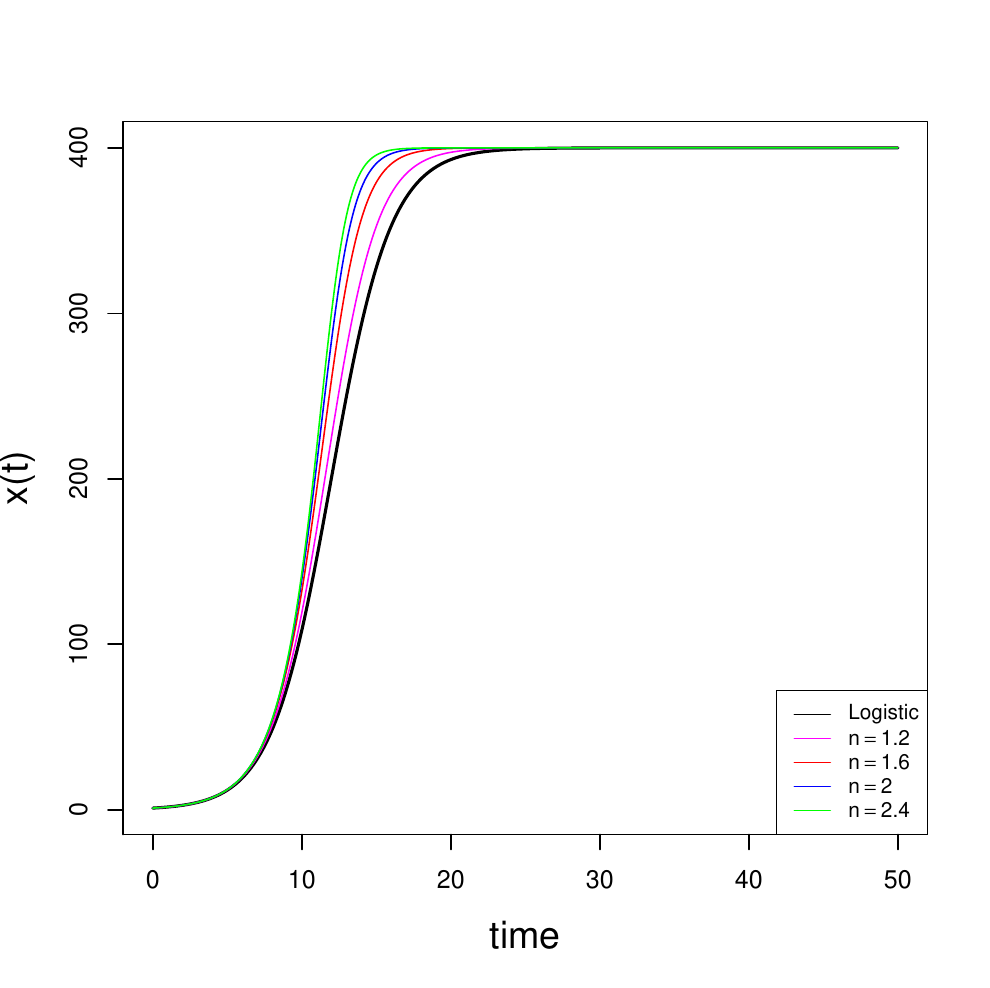}
\caption{Case 2: general curve in the case $p=1$ compared with the Logistic curve (black line), for several choices of $n$ ($<1$ on the left and $>1$ on the right).}\label{fig4}
\end{figure}

Case 2 in which $p=1$ is shown in Figure \ref{fig4} where the the general curve $x(t)$ in \eqref{sol1} is compared with the logistic curve, obtained by setting $p=1$ and $n=1$ as established in the Remark \ref{Remark2} for several choices of $n<1$ (on the left) and of $n>1$ (on the right). In both the cases, when $n$ increases, the curve $x(t)$ reaches its carrying capacity in a smaller time interval, so in the first plot the logic curve presents a growth velocity that is higher with respect to the others, while in the second plot, it is the curve milder. Further we note that for $n>1$ the curves seem closer to each ones, showing that the dependence on $n$ is not linear and it is more evident in the case $n<1$. In Table \ref{TABELLA} the information related to the inflection point $t_{Inf_1}:=t_{Inf}$ in the Case 2 is listed for $n=0.4, 0.6, 0.8$ and $1$, showing that also in this case $t_{Inf_1}$ decreases as $n$ increases, while the values  of $x(t_{Inf_1})$ and $\pi_x$ increase. \par
%
%
\begin{figure}[htbp]
\centering
\includegraphics[scale=0.35]{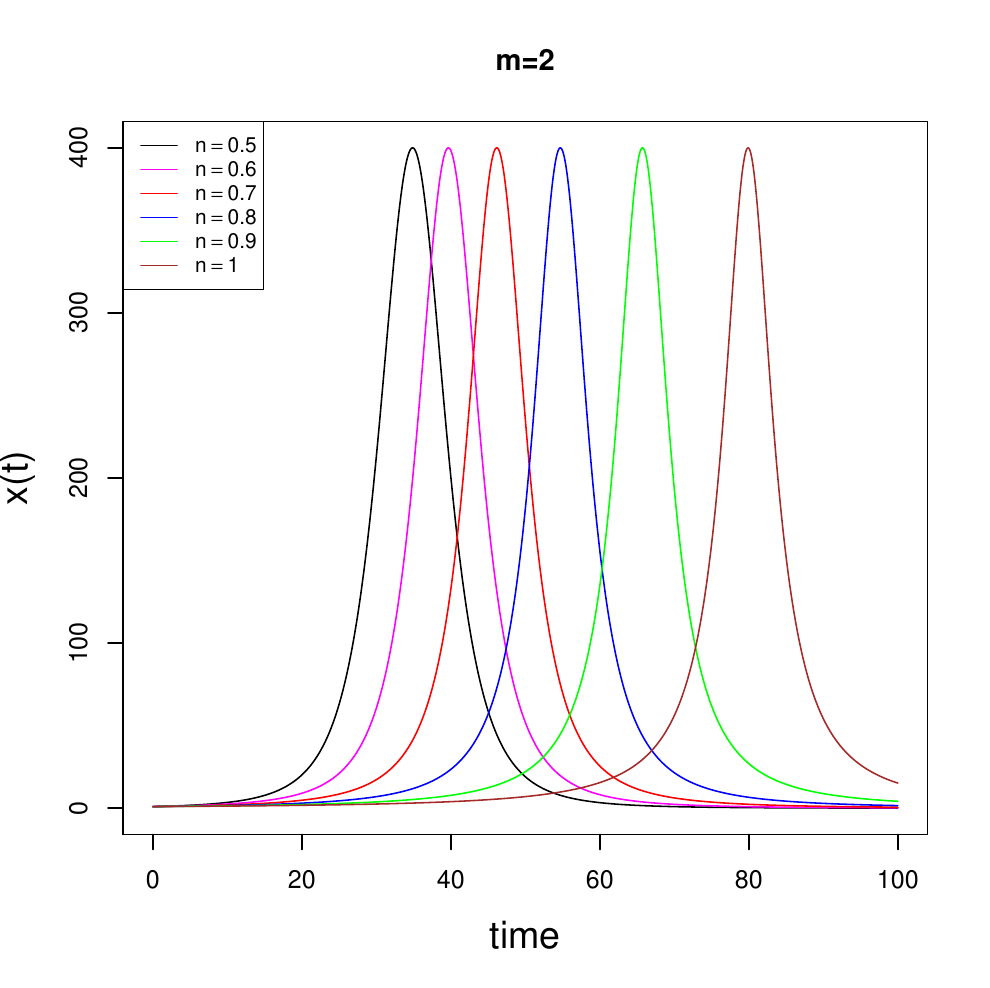}\includegraphics[scale=0.35]{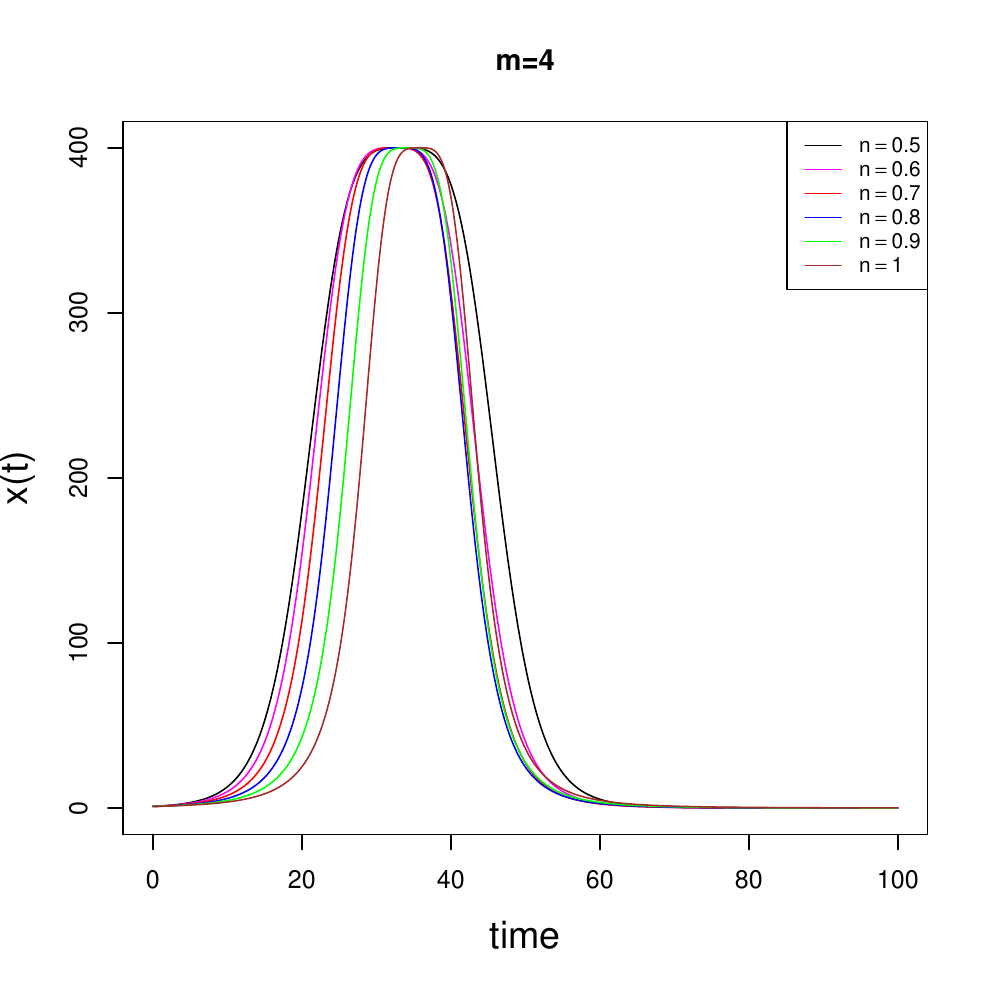}
\includegraphics[scale=0.35]{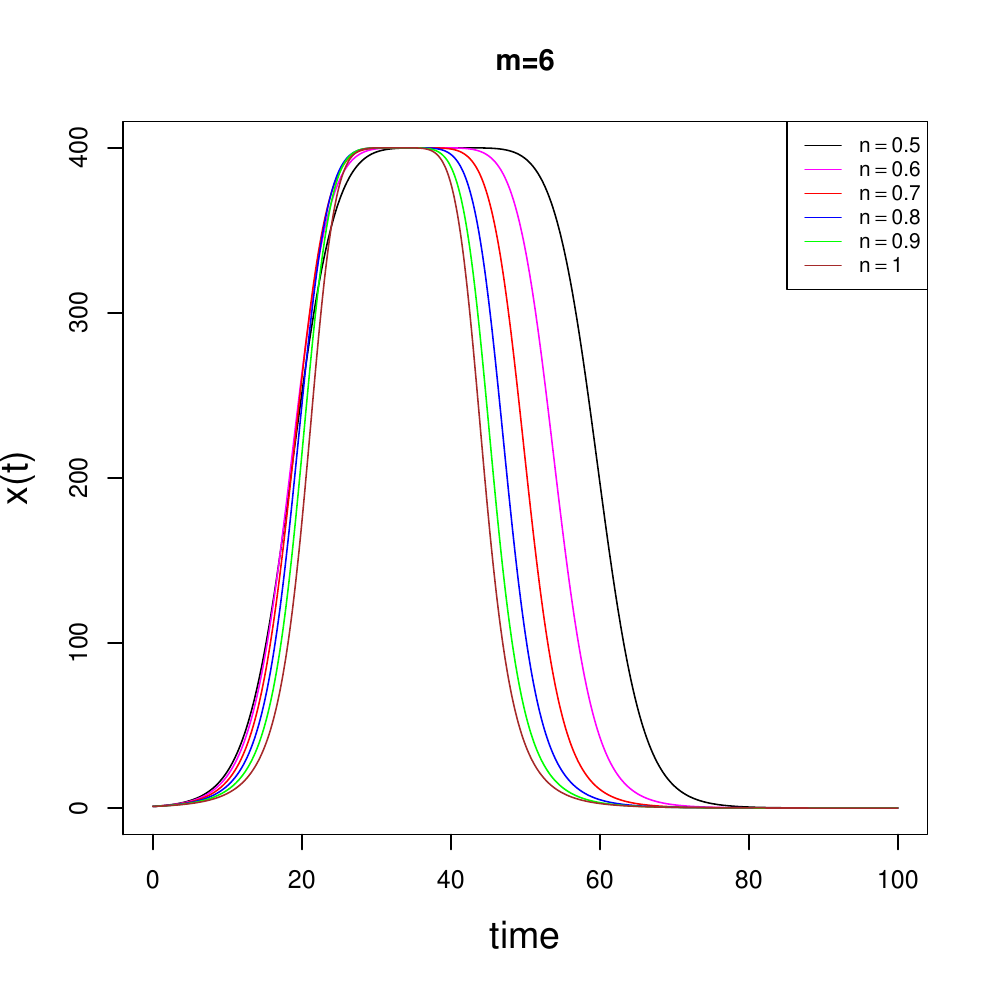}\includegraphics[scale=0.35]{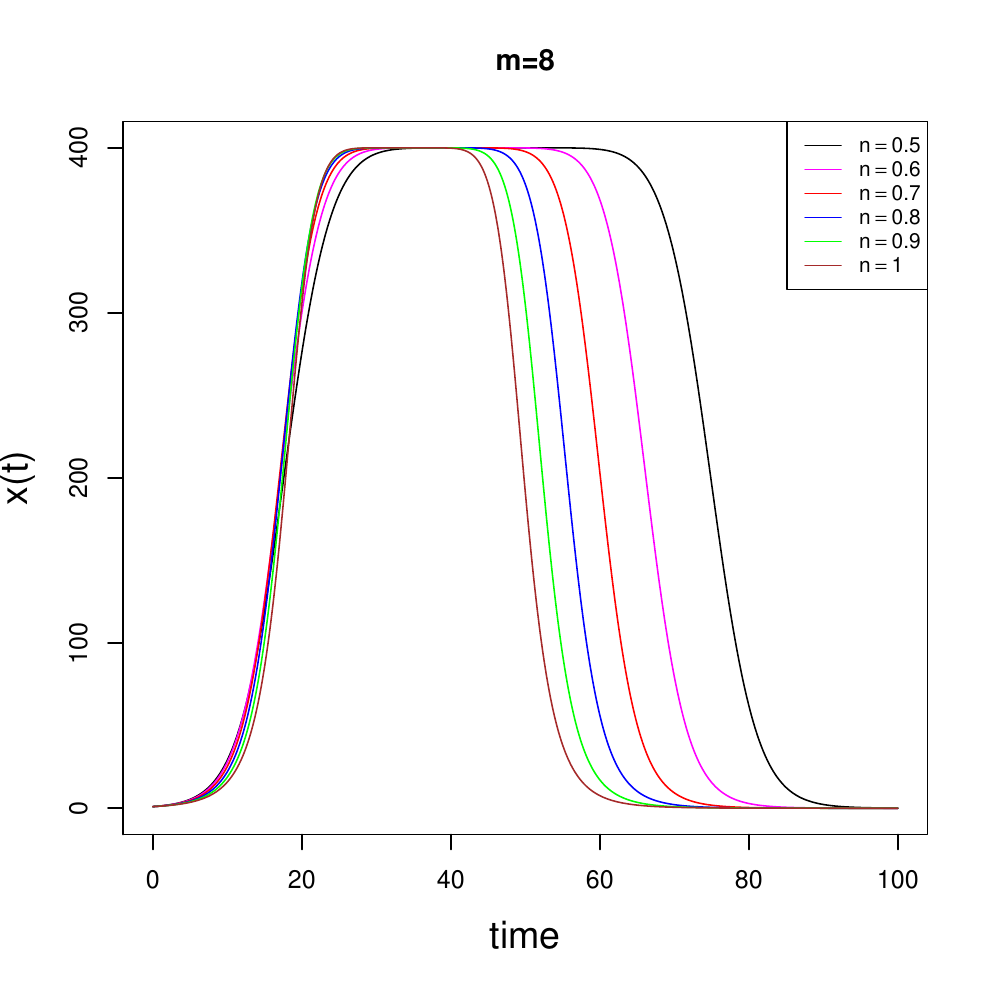}
\caption{Case 3a: general curve for several choices of $n$ and for $m=\frac{1}{1-p}=2,4,6$ and $8$ (from left to right and from top to bottom)}\label{fig5}
\end{figure}

\begin{figure}[htbp]
\centering
\includegraphics[scale=0.35]{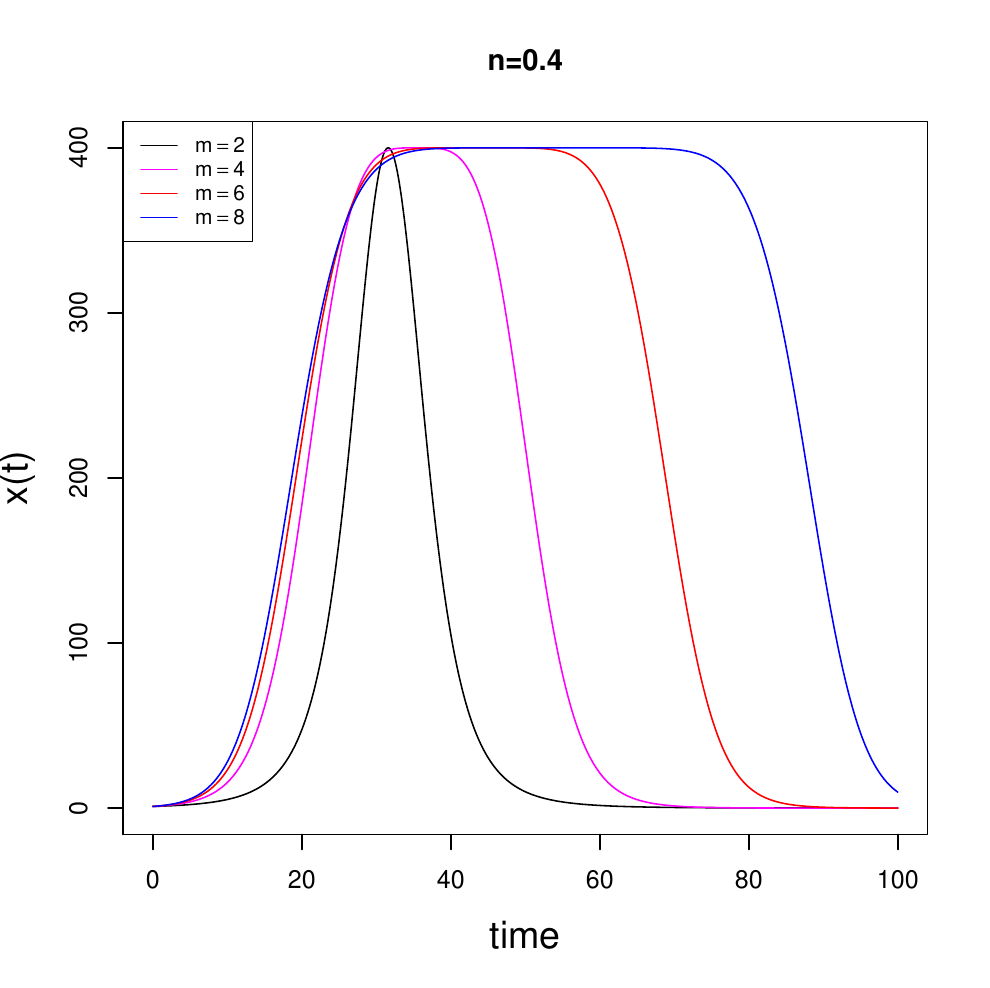}\includegraphics[scale=0.35]{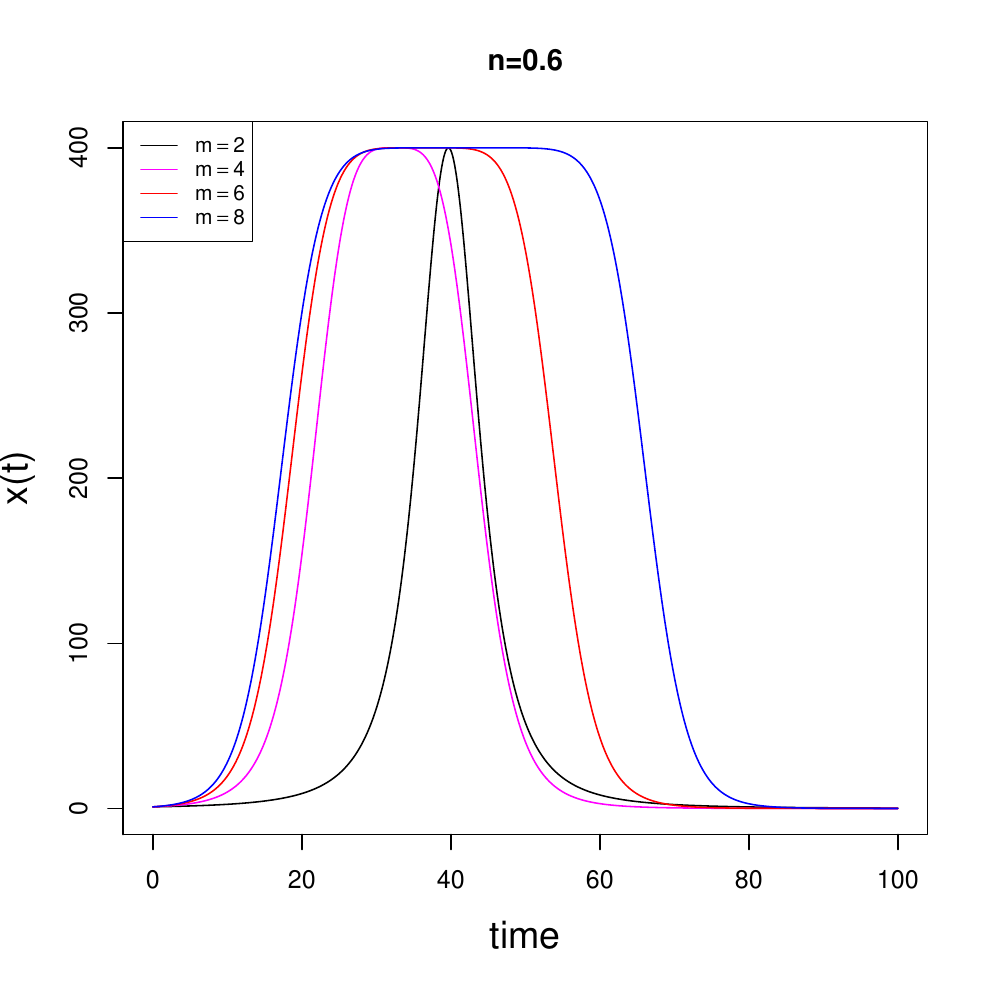} \\
\includegraphics[scale=0.35]{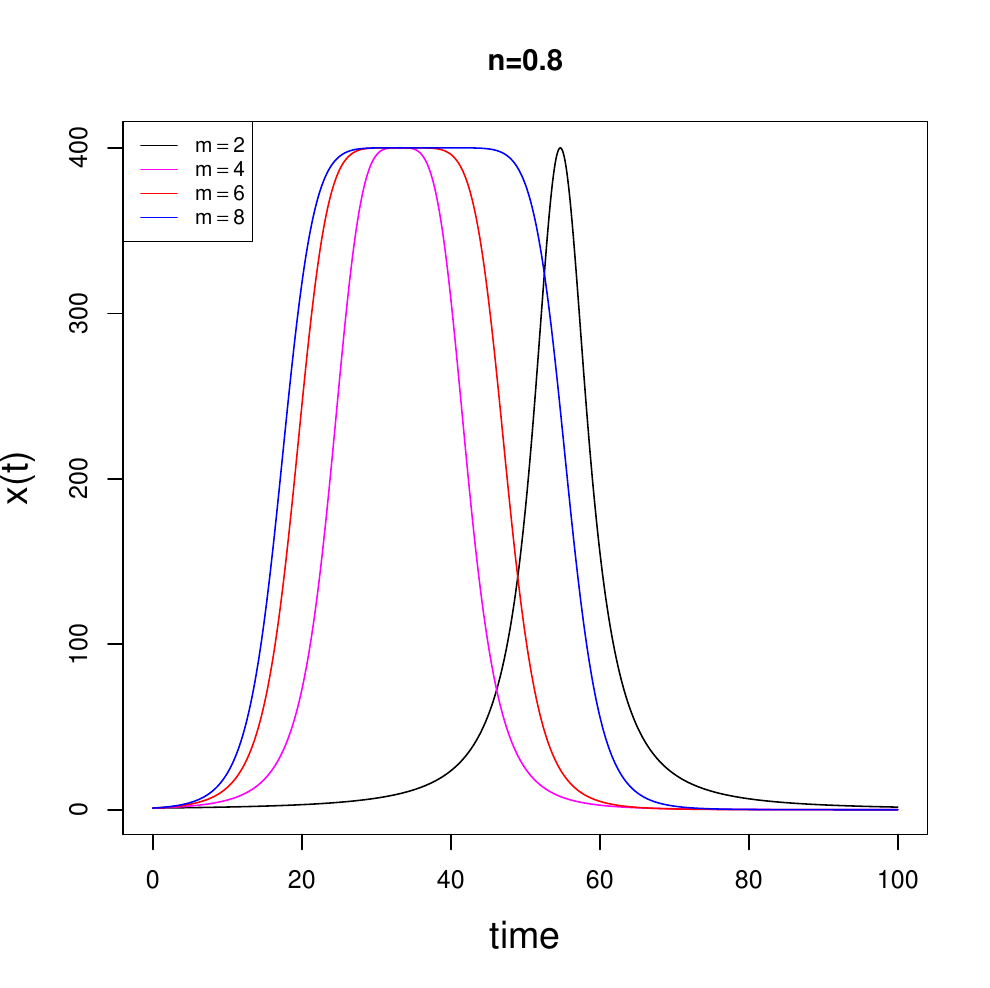}\includegraphics[scale=0.35]{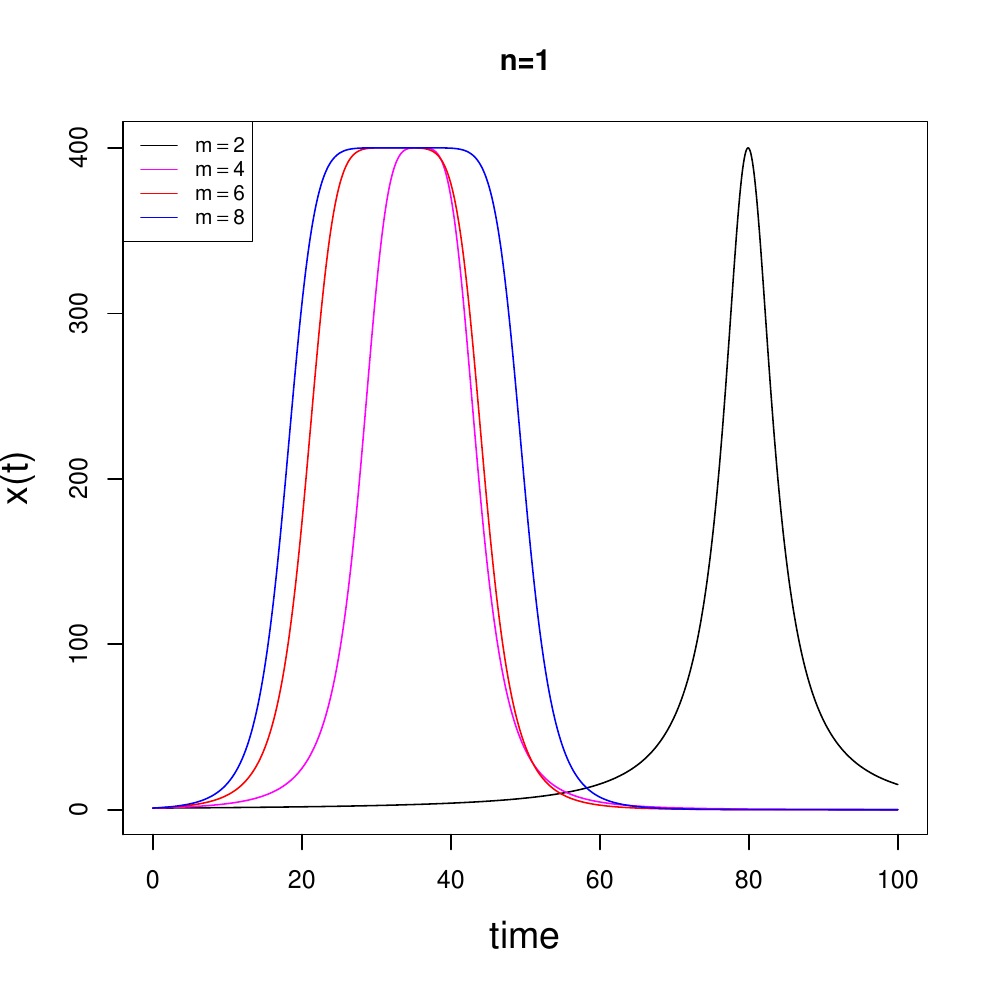}
\caption{Case 3a: general curve for $n=0.4, 0.6, 0.8, 1$ for several choices of the parameter $m=\frac{1}{1-p}=2,4,6$ and $8$.}\label{fig6}
\end{figure}

Let us consider now the Case 3 in which $0<p<1$. As pointed out in Section 3, in this case we need to consider three subcases:
\begin{enumerate}
      \item[a.] $\dfrac{1}{1-p}=m\in\mathbb{N}$, with $m$ even.
      \item[b.] $\dfrac{1}{1-p}=m\in\mathbb{N}$, with $m$ odd.
	  \item[c.] $\dfrac{1}{1-p}=m\notin\mathbb{N}$.
\end{enumerate}
In the Case 3a the curve $x(t)$ presents a maximum and it tends to $0$ for $t\to \infty$. Such case is illustrated in Figures \ref{fig5} and \ref{fig6}. In Figure \ref{fig5} the general curve is illustrated for $m=2,4,6$ and $8$ (from the top to the bottom), i.e. $p=\frac{1}{2}, \frac{3}{4}, \frac{5}{6}, \frac{7}{8}$, respectively. In each plot several choices of the parameter $n$ are considered ranking from $0.5$ to 1. The behavior of the curve is quite unexpected, since in the case $p=\frac{1}{2}$ it seems that the role of $n$ is to  translate ahead the curve as $n$ increases. In the other cases, i.e. for $p=\frac{3}{4}, \frac{5}{6}, \frac{7}{8}$, we observe a \lq\lq plateau\rq\rq\  that is wider as $p$ increases. Anyway we point out that the maximum of the curve $x(t)$ in \eqref{sol1} is always unique and with coordinates $(t^*,k)$ with $t^*$ in \eqref{tStar}, contrary to what it seems from the plots. Figure \ref{fig6} shows the behavior of the curve $x(t)$ for $n=0.4,0.6,0.8$ and $1$ (from the top to the bottom). Here it is evident the peculiar behavior of the curve in which $m=2$, i.e. $p=\frac{1}{2}$ that is translated with respect to the other curves, while the main change for the cases $m=4,6,8$ is the decrease of the width of the plateau as the parameter $n$ increases. Clearly, in this case the inflection points are two and the results related to them are illustrated in \ref{TABELLA}, in which $x(t_{Inf_1})=x(t_{Inf_2})$. The inflection point $t_{Inf_1}$ decreases as $p$ increases, while $t_{Inf_2}$ increases. A less regular behavior is observed for $t_{Inf_1}$ and $t_{Inf_2}$ as $n$ changes. Further, $x(t_{Inf_1})$ decreases as $p$ increases, while it increases as $n$ increases. \par
%
%
%
%
%
\begin{figure}[htbp]
\centering
\includegraphics[scale=0.35]{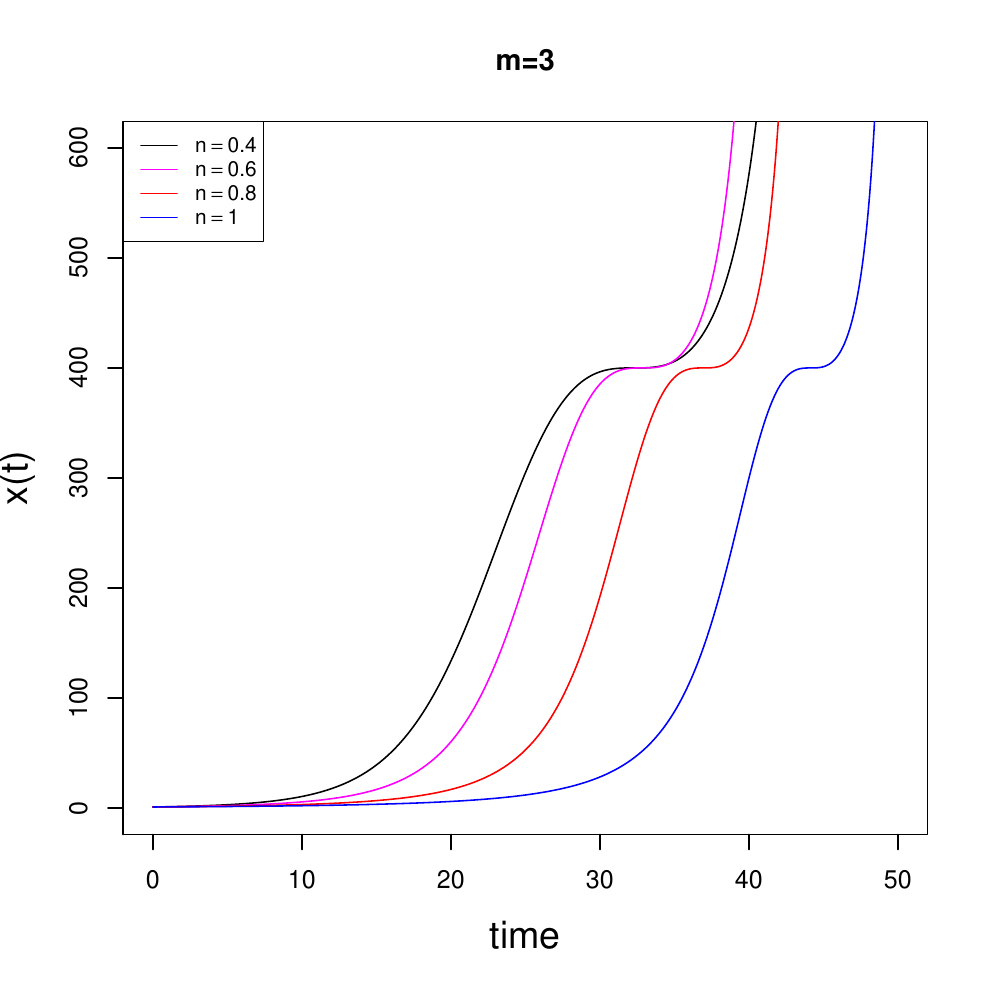}\includegraphics[scale=0.35]{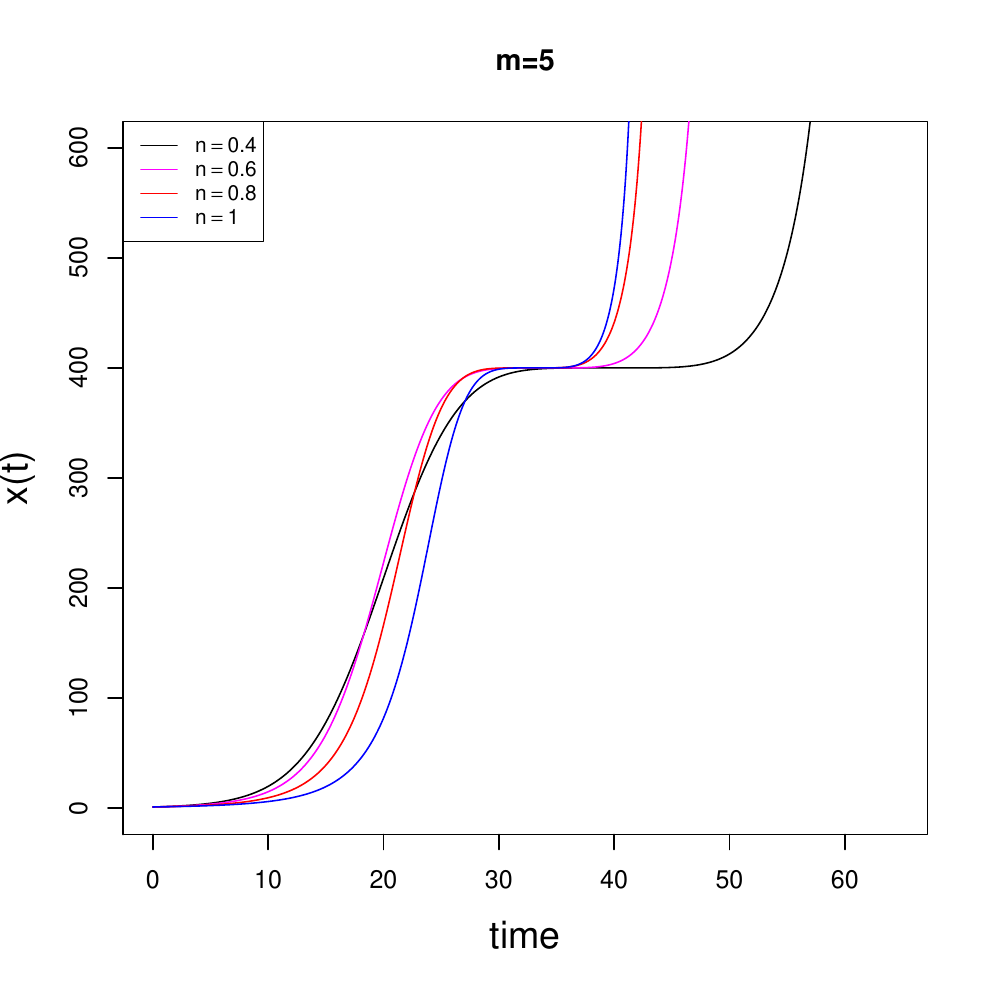}\\
\includegraphics[scale=0.35]{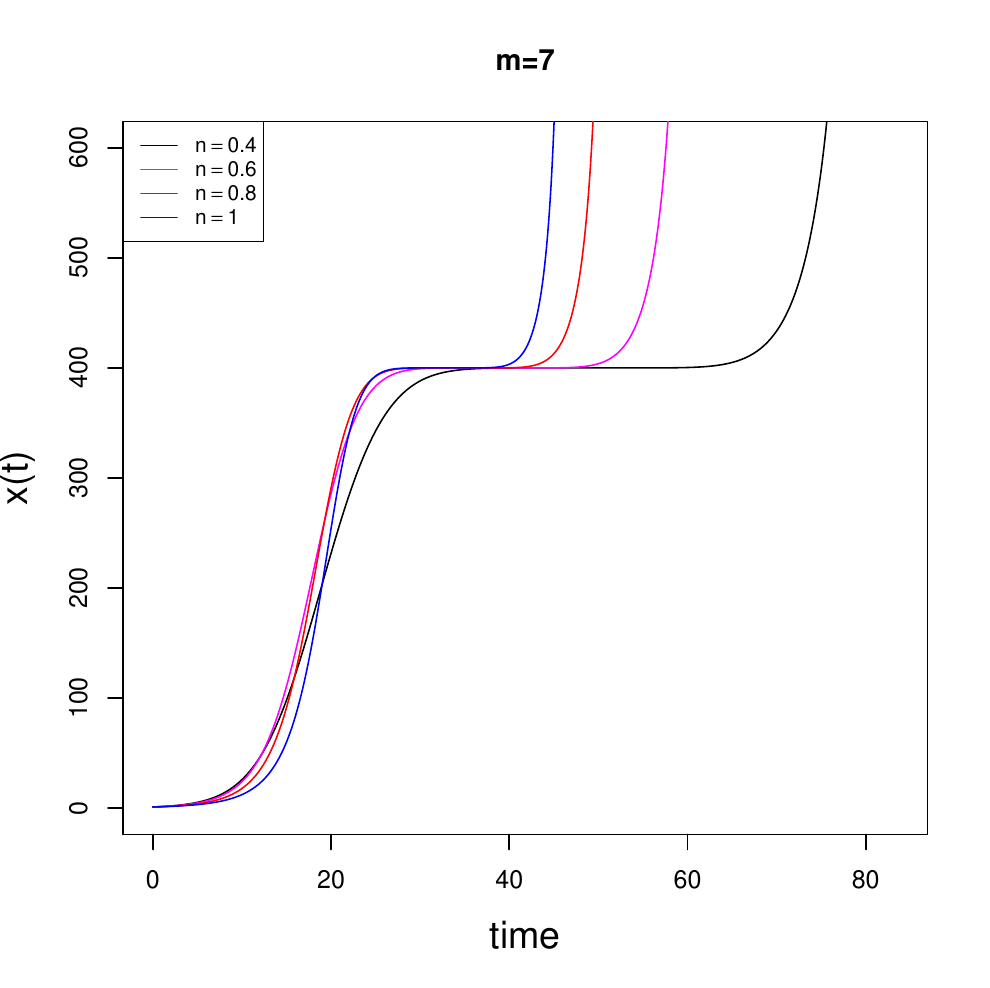}\includegraphics[scale=0.35]{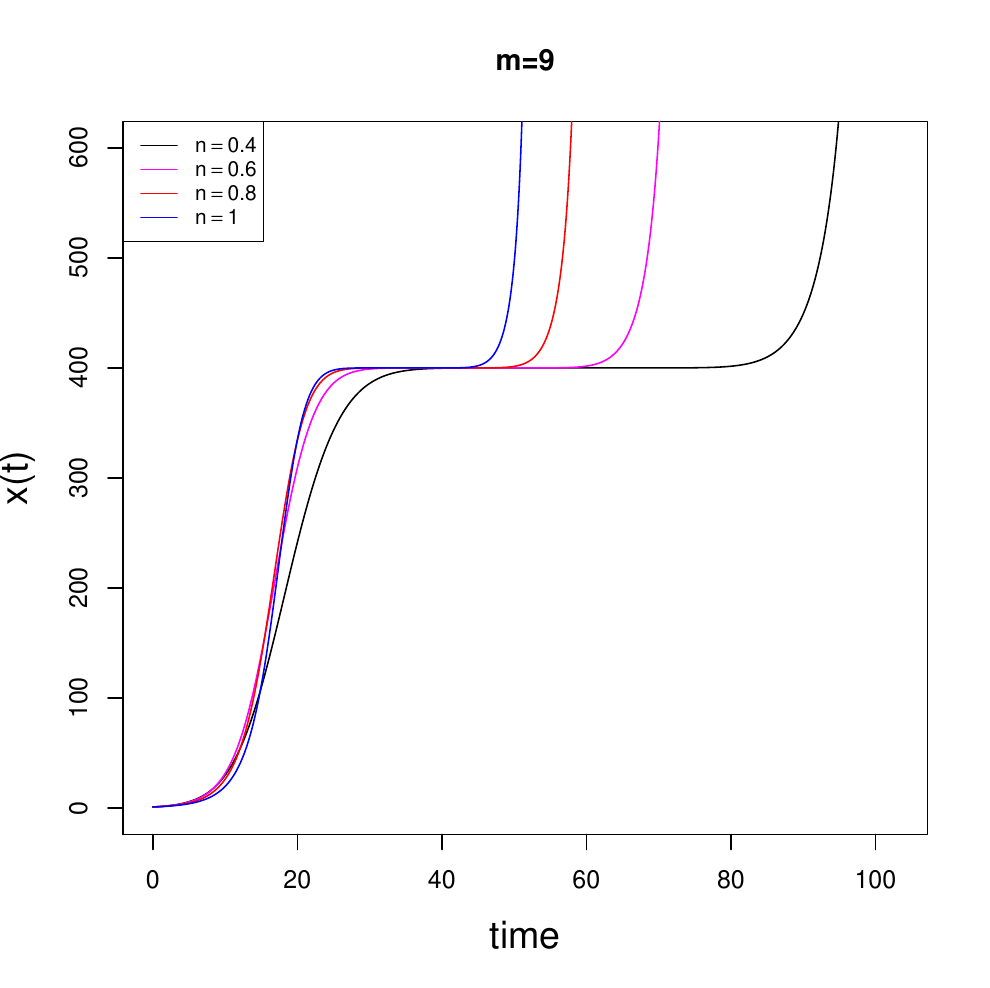}
\caption{Case 3b: general curve for several choices of $n$ and for $m=\frac{1}{1-p}=3,5,7$ and $9$ (from left to right and from top to bottom).}\label{fig7}
\end{figure}

\begin{figure}[htbp]
\centering
\includegraphics[scale=0.35]{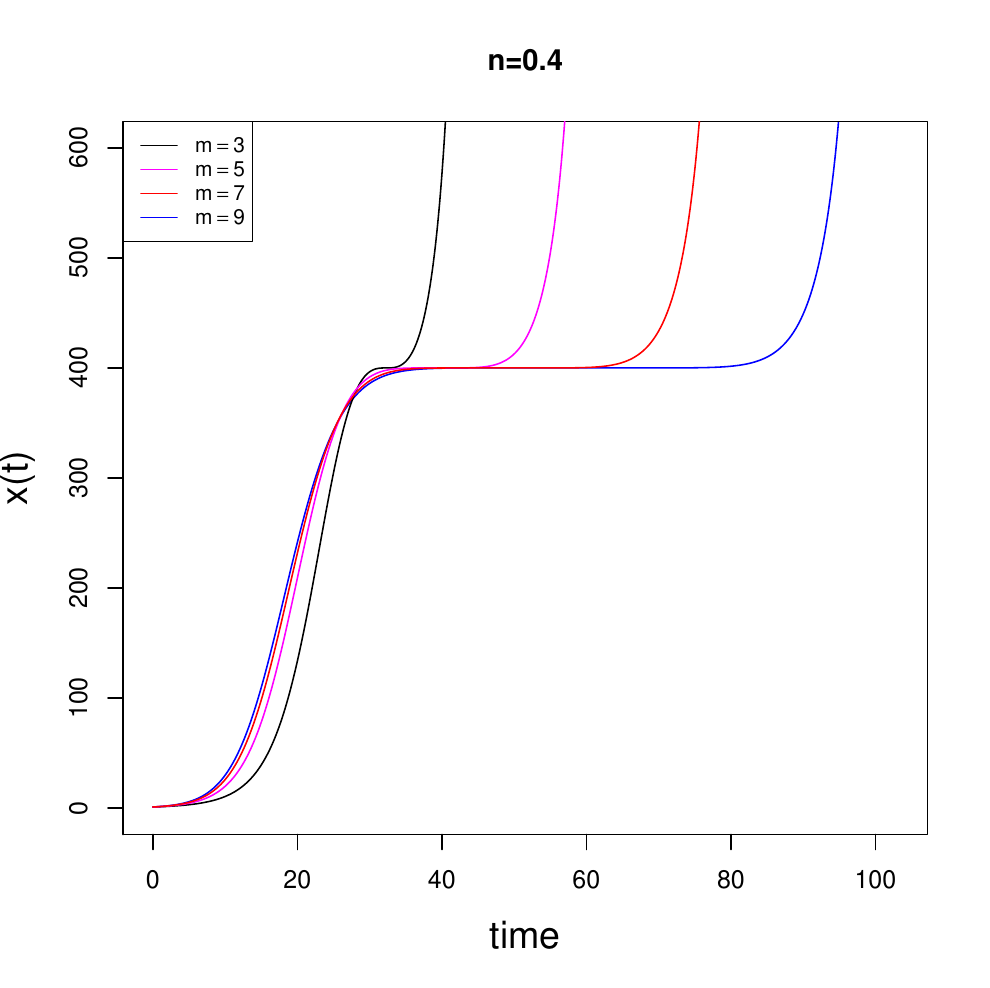}\includegraphics[scale=0.35]{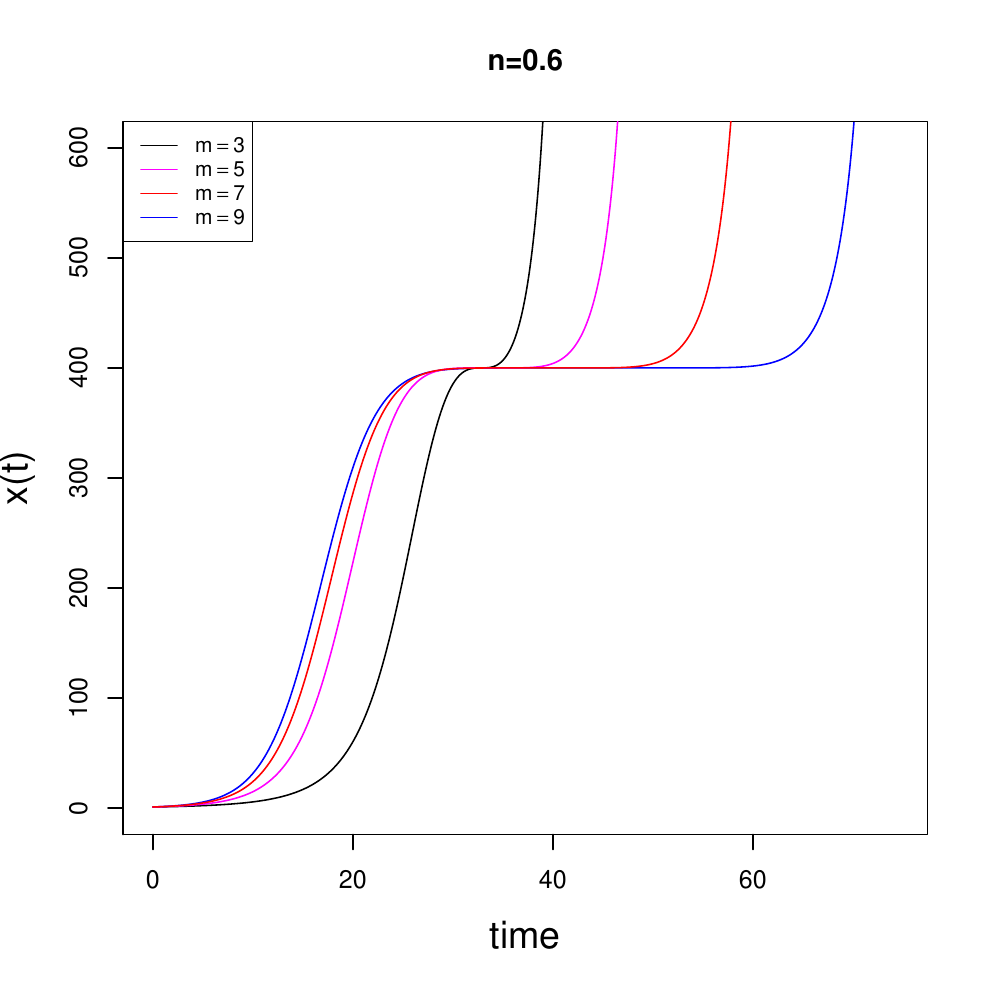}\\
\includegraphics[scale=0.35]{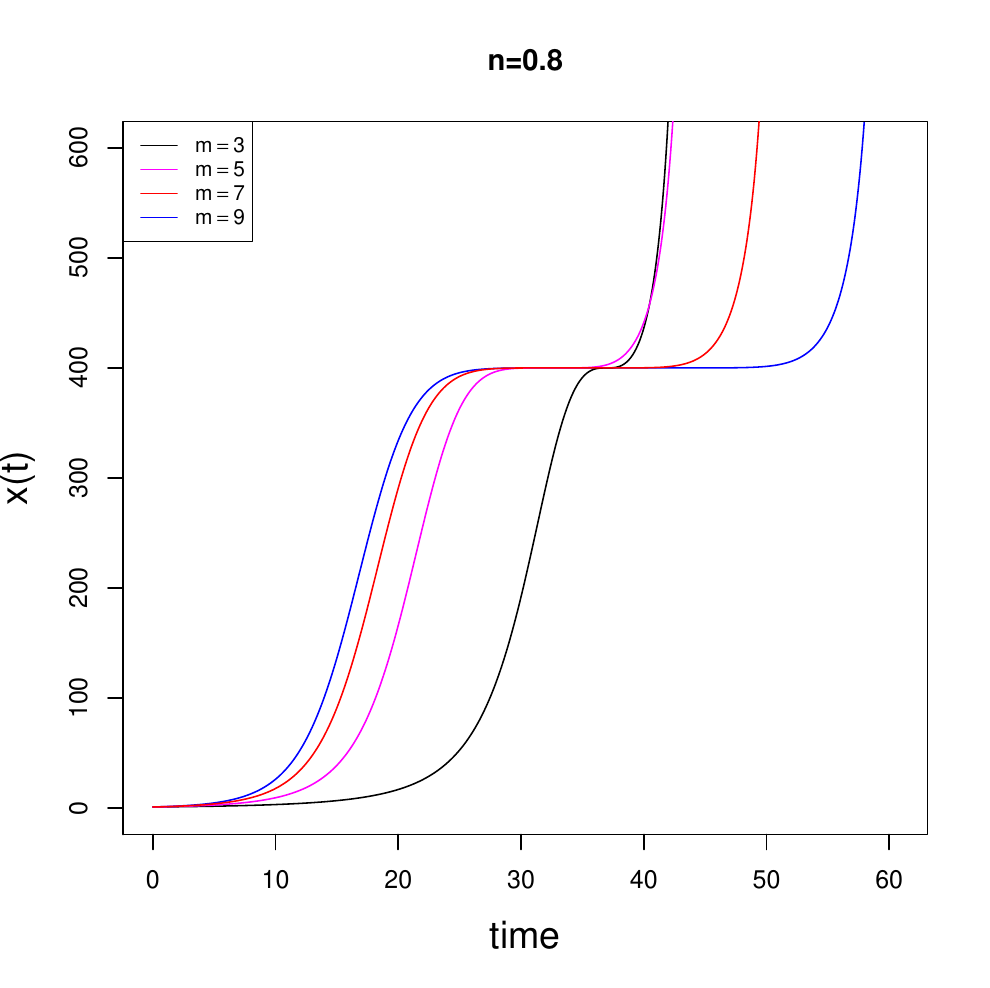}\includegraphics[scale=0.35]{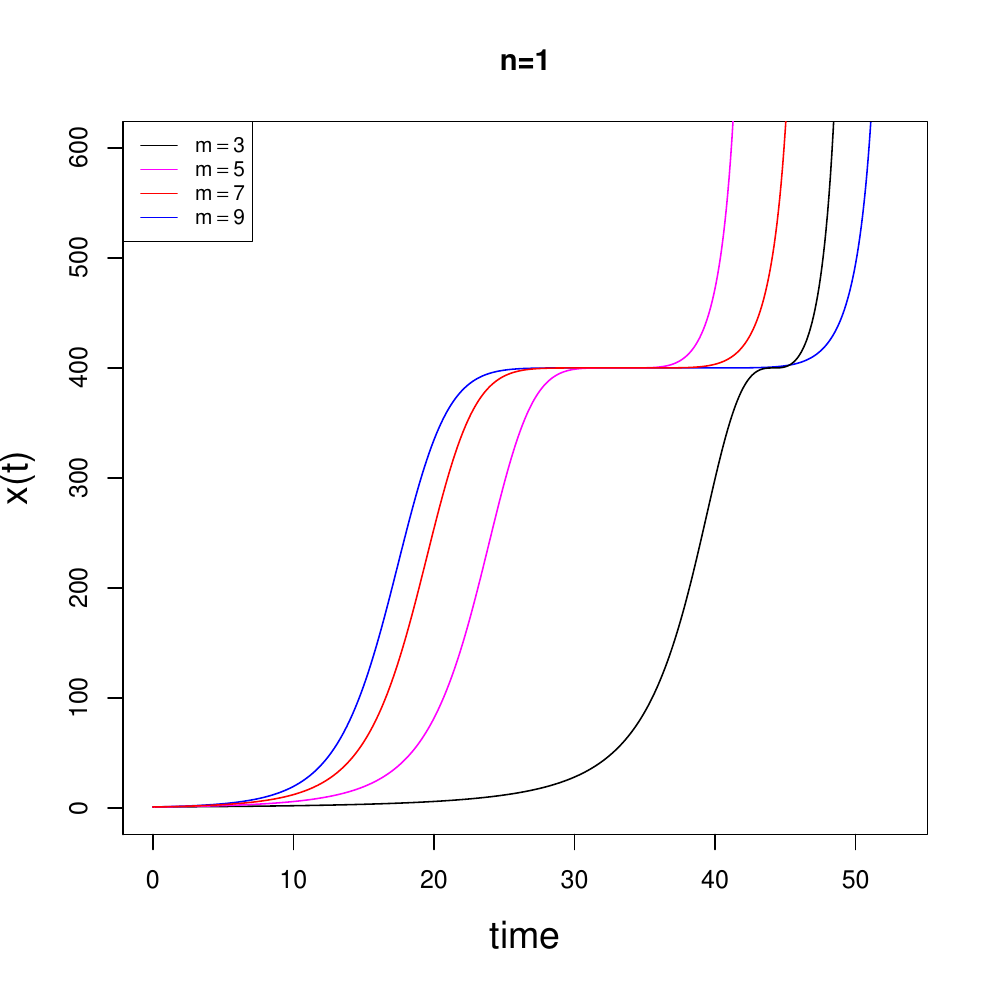}
\caption{Case 3b: general curve for $n=0.4, 0.6, 0.8,1$ for several choices of the parameter $m=\frac{1}{1-p}=3,5,7$ and $9$. }\label{fig8}
\end{figure}

In the Case 3b the curve $x(t)$ is defined in $[t_0,t_1)$ with $t_1$ given in \eqref{t1} and $\lim_{t\to t_1} x(t)=\infty$, so the population explodes in a finite time interval. The results of this case are shown in Figures \ref{fig7} and \ref{fig8}. In Figure \ref{fig7} the curve $x(t)$ is plotted for $m=3,5,7$ and $9$ (from the top to the bottom) and for $n=0.4, 0.6, 0.8$ and $1$. In all the plots the curves present a growth velocity depending on the parameter $n$, in particular in the first subinterval in which $x(t)<k$. Further, $x(t)$ presents a plateau near $k=400$, which width increases as $p$ increases. In this case the value $k$ doesn't correspond to the maximum of the function $x(t)$, since after the plateau the curve increases indefinitely. Indeed, as pointed in \eqref{Caso3b}, $k=400$ is the value of the function in $t_{Inf_2}$. Figure \ref{fig8}, in which $x(t)$ is plotted for $n=0.4, 0.6, 0.8$ and $1$ (from the top to the bottom) and for $m=3,5,7$ and $9$, confirms the results observed in Figure \ref{fig7}. In Table \ref{TABELLA} the values of $t_{Inf_1}$ and $t_{Inf_2}$, $x(t_{Inf_1})$ and the proportion $\pi_x=k/x(t_{Inf_1})$ are shown. Clearly the proportion $k/x(t_{Inf_2})=1$ since $x(t_{Inf_2})=k$. We observe that $t_{Inf_1}$ decreases as $p$ increases and it increases as $n$ increases; $t_{Inf_2}$ presents a more irregular behavior. The values of $x(t_{Inf_1})$ and $\pi_x$ decrease as $p$ increases and increase for increasing $n$. \par
%
%
%
%
%
%
\begin{figure}[htbp]
\centering
\includegraphics[scale=0.35]{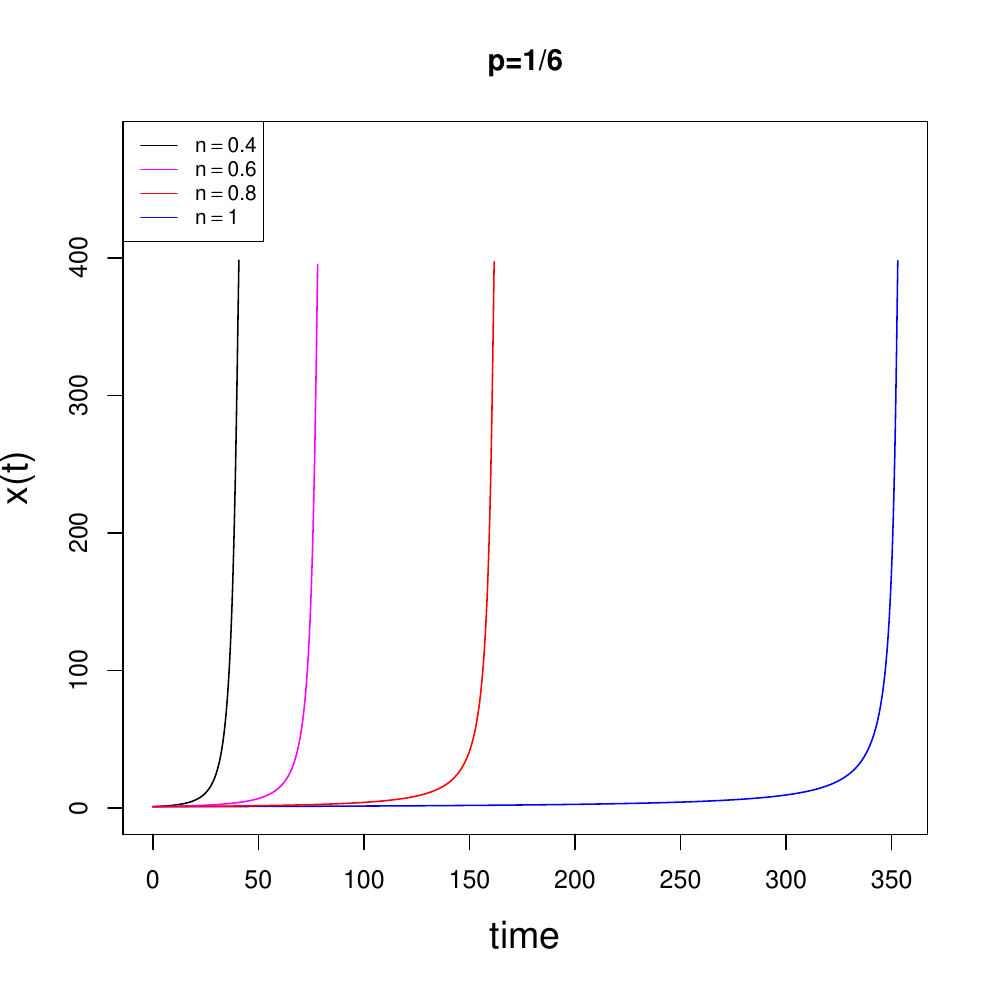}\includegraphics[scale=0.35]{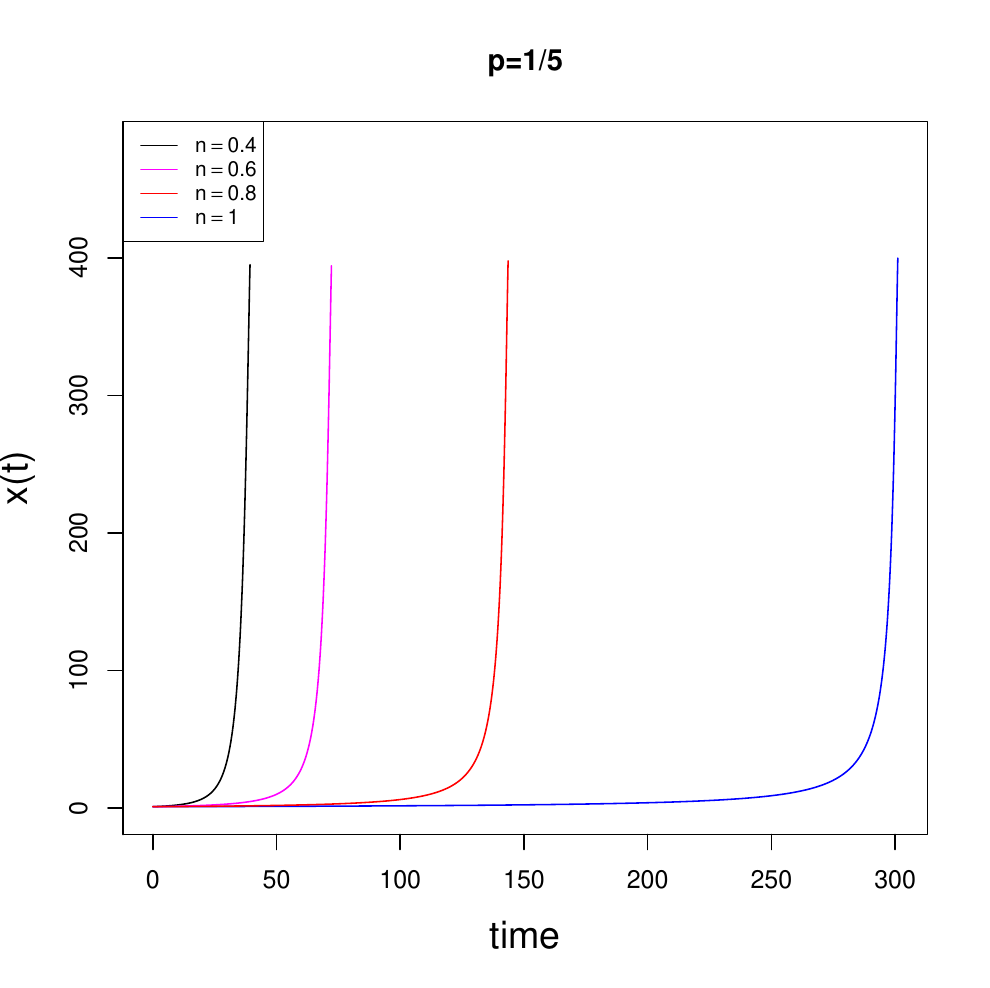}\\
\includegraphics[scale=0.35]{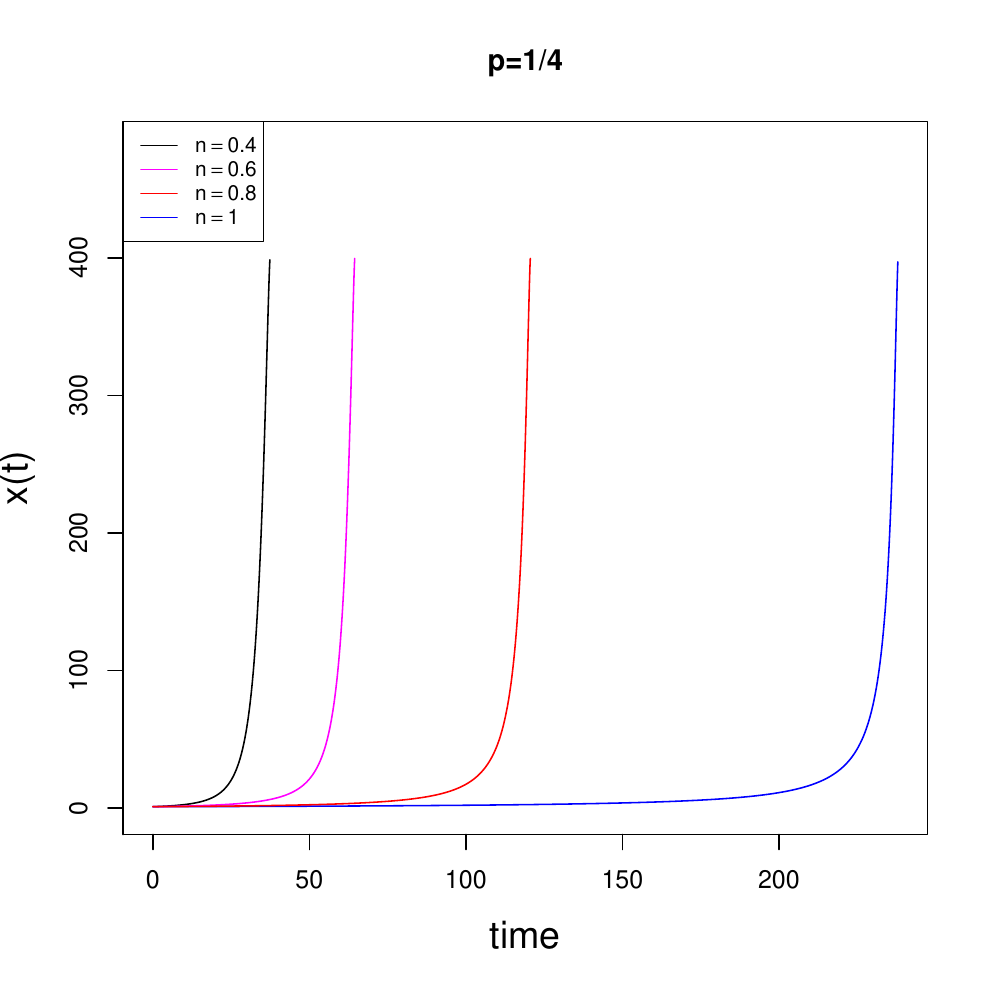}\includegraphics[scale=0.35]{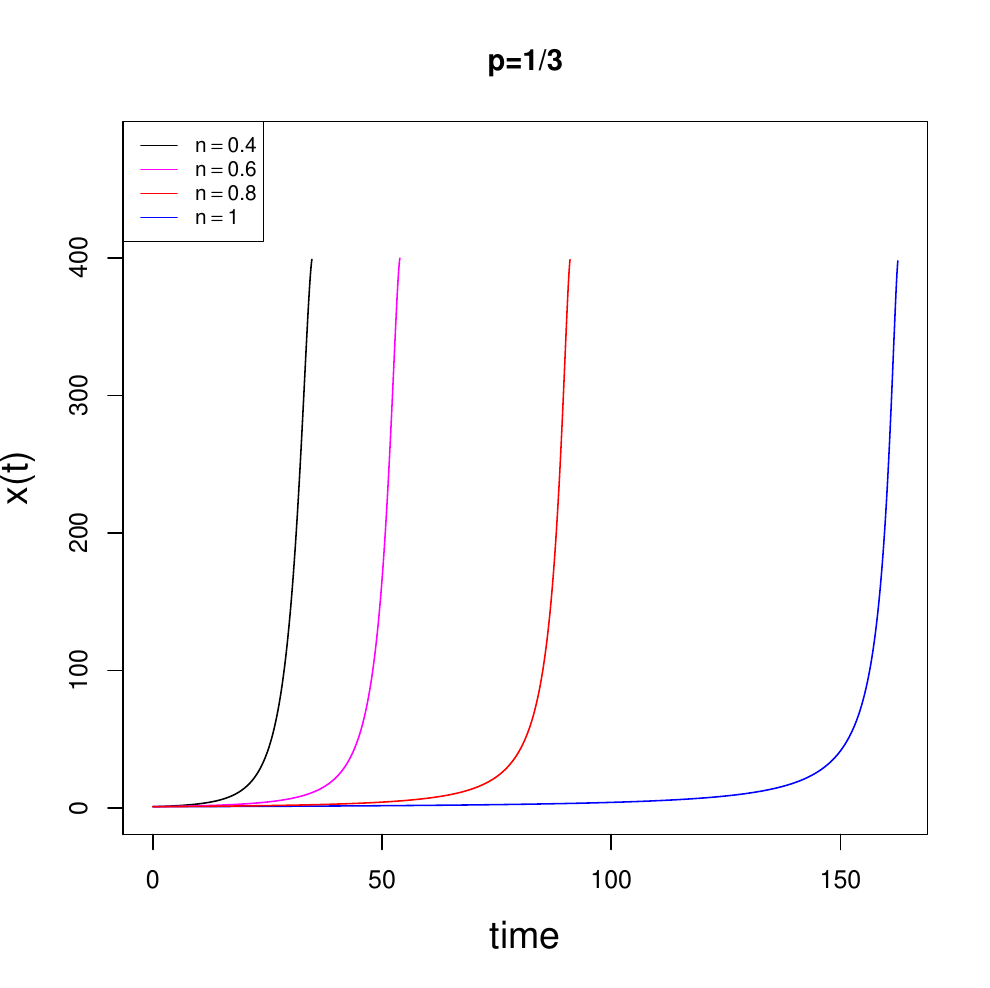}
\caption{Case 3c: general curve for several choices of $n$ and for $p=\frac{1}{6}, \frac{1}{5}, \frac{1}{4}$ and $\frac{1}{3}$ (from left to right and from top to bottom).}\label{fig9}
\end{figure}

\begin{figure}[htbp]
\centering
\includegraphics[scale=0.35]{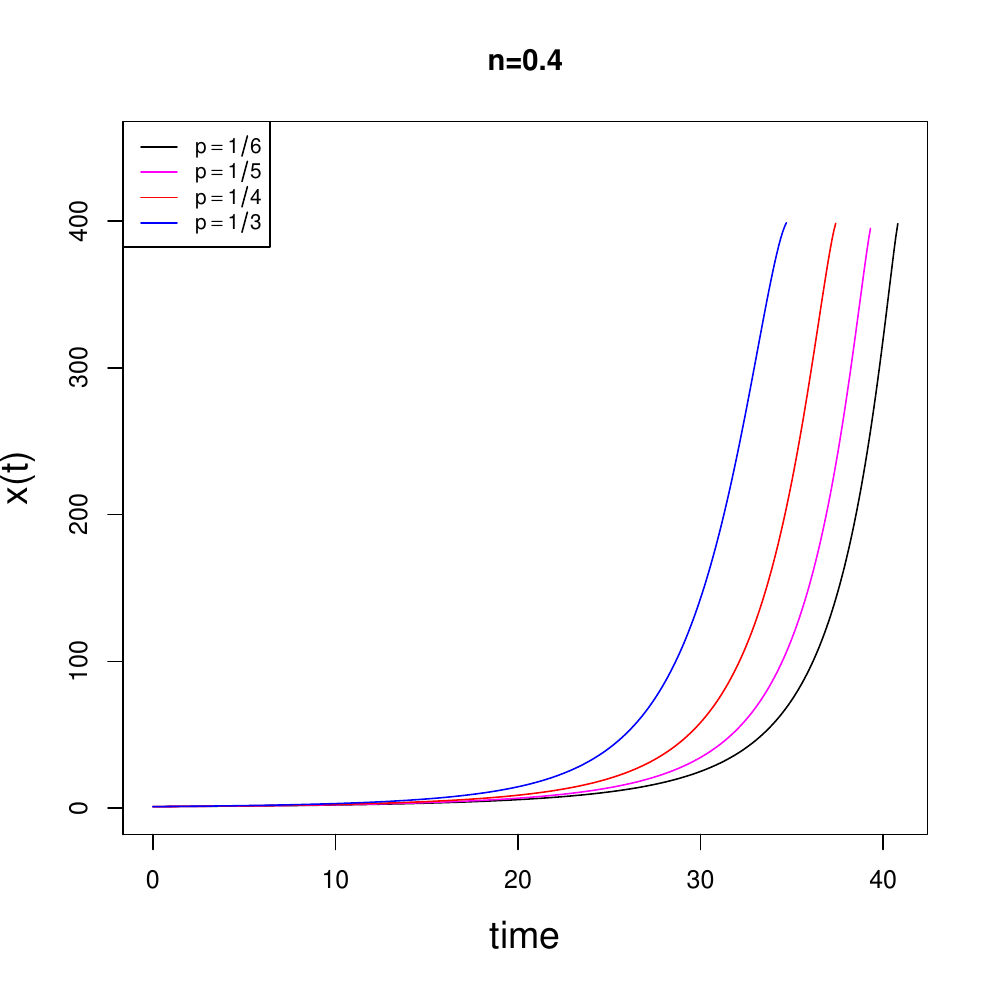}\includegraphics[scale=0.35]{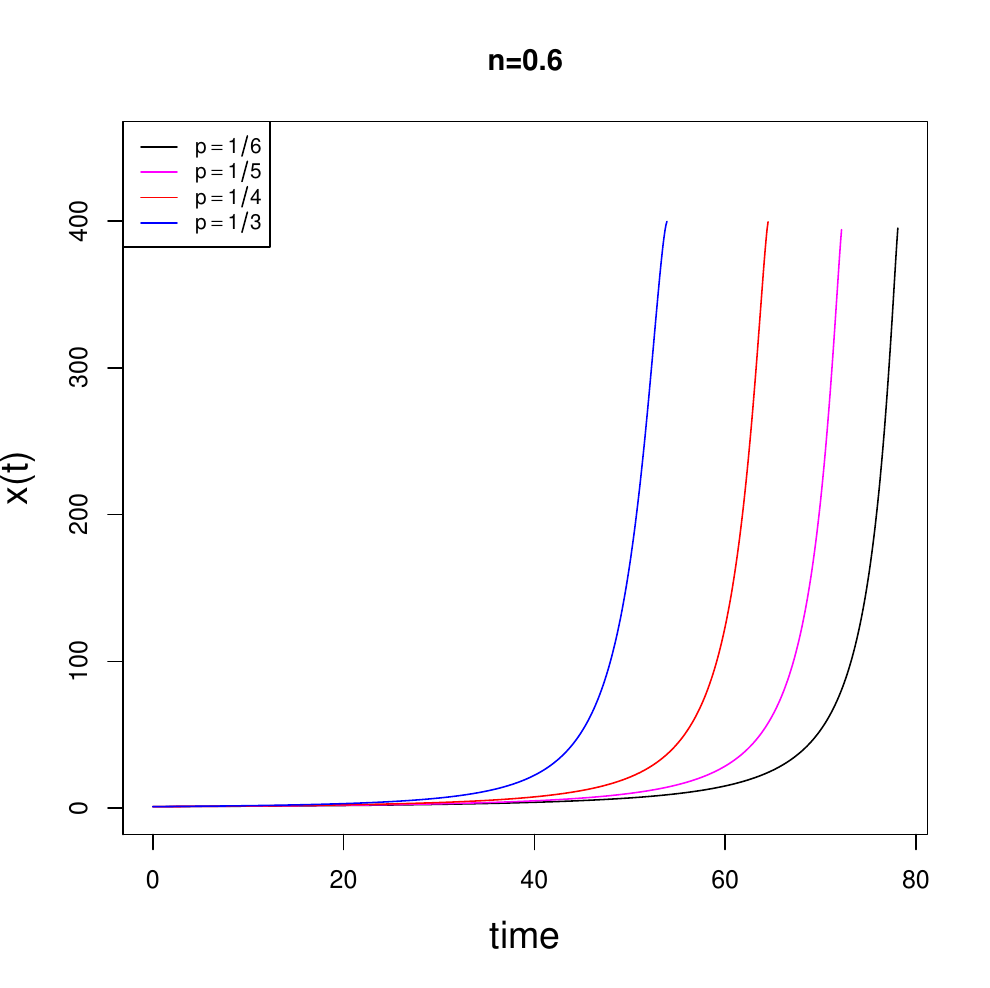}\\
\includegraphics[scale=0.35]{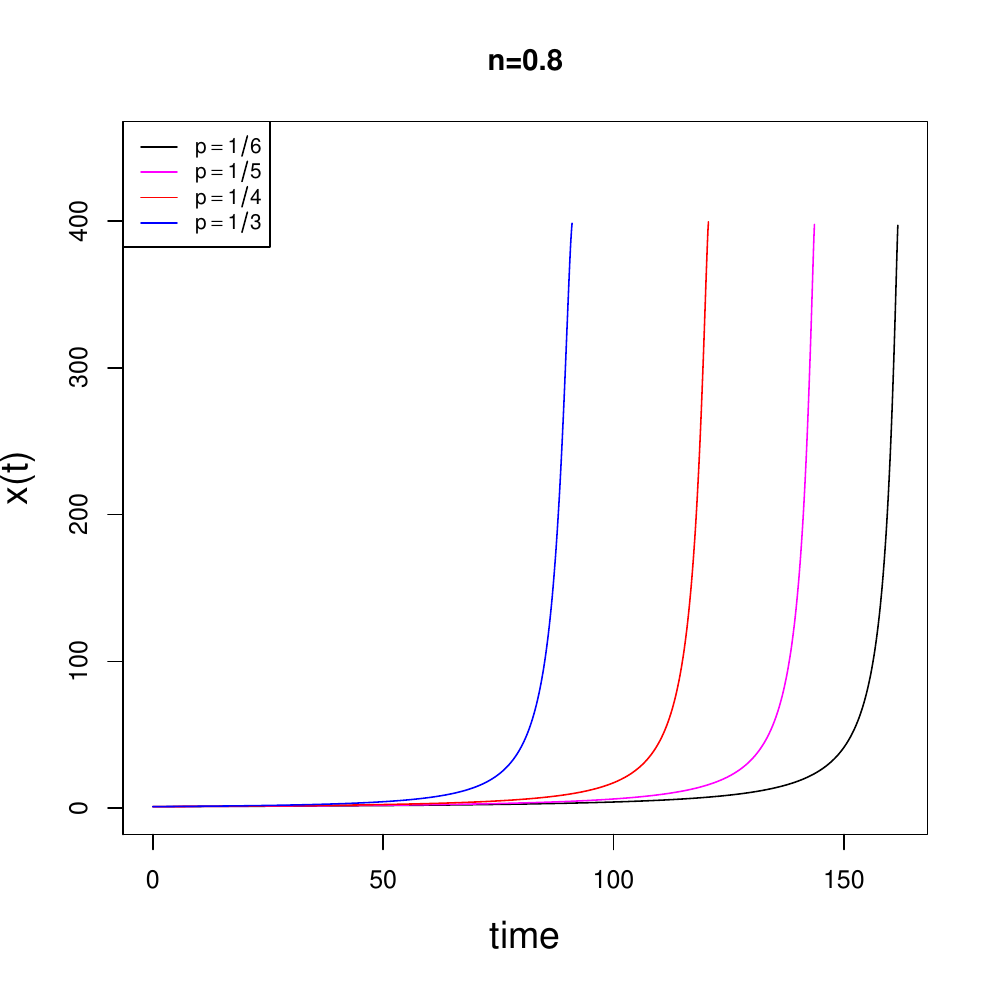}\includegraphics[scale=0.35]{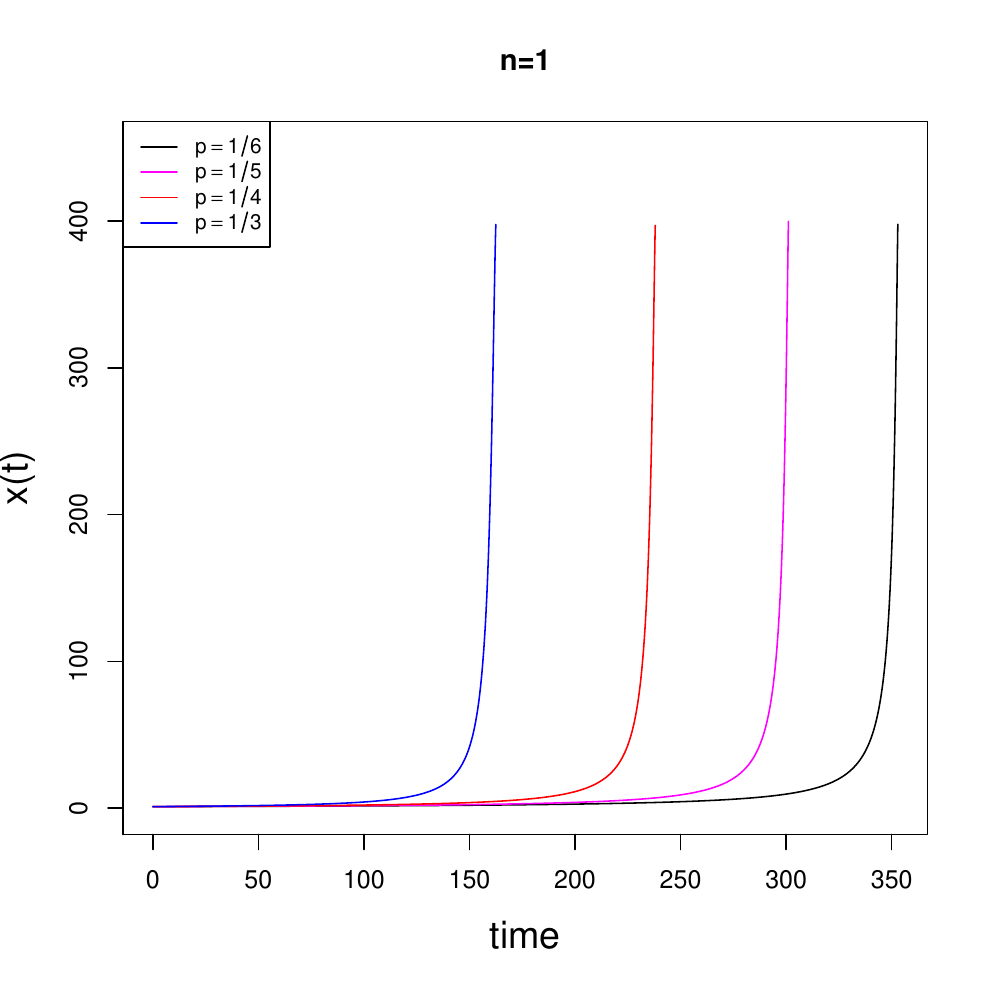}
\caption{Case 3c: general curve for $n=0.4, 0.6, 0.8,1$ for several choices of the parameter $p=\frac{1}{6}, \frac{1}{5}, \frac{1}{4}$ and $\frac{1}{3}$. }\label{fig10}
\end{figure}

In the Case 3c the curve $x(t)$ is defined in $[t_0,t_2)$ with $t_2$ given in \eqref{t2} and $\lim_{t\to t_2} x(t)=k$, so the population reaches its carrying capacity at a finite time interval. In Figure \ref{fig9} the curve $x(t)$ is plotted for $p=1/6, 1/5, 1/4$ and $1/3$ (from the top to the bottom) and for $n=0.4, 0.6, 0.8$ and $1$. In all the plots it is evident that the population size remains near to 0 for a time interval which amplitude increases as $n$ increases depends increases as $p$ increases. In Figure \ref{fig10}, $x(t)$ is plotted for $n=0.4, 0.6, 0.8$ and $1$ (from the top to the bottom) and for $p=1/6, 1/5, 1/4$ and $1/3$; we note that in this case $x(t)$ is near 0 for a time interval that is smaller as $p$ increases.  Table \ref{TABELLA} confirms the results of Figures \ref{fig9} and \ref{fig10} showing that the values of $t_{Inf_1}:=t_{Inf}$ and $x(t_{Inf_1})$ decrease for increasing  $p$, while they increase as $n$ increases; the proportion $\pi_x=k/x(t_{Inf_1})$ follows the same behavior of $x(t_{Inf_1})$.
%
%
%
%
%
%
%
\newpage
%
\begin{table}[htbp]
\begin{center}
\begin{sideways}
{\tiny
\begin{tabular}{|c|cccc|cccc|cccc|cccc|}
\hline\hline
\multicolumn{17}{|c|}{Case 1: $1<p<1+\frac{1}{n}$}\\
\hline
&\multicolumn{4}{|c|}{$n=0.4$}&\multicolumn{4}{|c|}{$n=0.6$}&\multicolumn{4}{|c|}{$n=0.8$}&\multicolumn{4}{|c|}{$n=1$}\\
\hline
$p$  & $t_{Inf_1}$ & $t_{Inf_2}$ & $x(t_{Inf_1})$ & $\pi_x$ & $t_{Inf_1}$ & $t_{Inf_2}$ & $x(t_{Inf_1})$ & $\pi_x$ & $t_{Inf_1}$ & $t_{Inf_2}$ & $x(t_{Inf_1})$ & $\pi_x$ & $t_{Inf_1}$ & $t_{Inf_2}$ & $x(t_{Inf_1})$ & $\pi_x$ \\
\hline
1.1 & 14.3349 & -- & 155.7464 & 0.3894 & 11.2005 & -- & 164.844 & 0.4121 & 9.6183 & -- & 172.8638 & 0.4322 & 8.6143 & -- & 180 & 0.45\\
1.2 & 12.6956 & -- & 140.0263 & 0.3501 & 9.183 & -- & 147.6832 & 0.3692 & 7.3699 & -- & 154.2831 & 0.3857 & 6.2025 & -- & 160 & 0.4 \\
1.3 & 11.1593 & -- & 125.2987 & 0.3132 & 7.4683 & -- & 131.2854 & 0.3282 & 5.6016 & -- & 136.1402 & 0.3404 & 4.4311 & -- & 140 & 0.35\\
1.4 & 9.7218 & -- & 111.5419 & 0.2789 & 6.0057 & -- & 115.6687 & 0.2892 & 4.1973 & -- & 118.4692 & 0.2962 & 3.1071 & -- & 120 & 0.3\\
1.5 & 8.3824 & -- & 98.7336 & 0.2468 & 4.7591 & -- & 100.8533 & 0.2521 & 3.0785 & -- & 101.3114 &  0.2533 & 2.1092 & -- & 100 & 0.25\\
1.6 & 7.1434 & -- & 86.8508 & 0.2171 & 3.7024 & -- & 86.8614 & 0.2172 & 2.1909 & -- & 84.7176 & 0.2118 & 1.3592 & -- & 80 &  0.2\\
1.7 & 6.0078 & -- & 75.8702 & 0.1897 & 2.8154 & -- & 73.718 & 0.1843 & 1.4956 & -- & 68.752 & 0.1719 & 0.8052 & -- & 60 & 0.15\\
1.8 & 4.9789 & -- & 65.7676 & 0.1644 & 2.0818 & -- & 61.4515 & 0.1536 & 0.9629 & -- & 53.4992 & 0.1337 & 0.4103 & -- & 40 & 0.1\\
1.9 & 4.0591 & -- & 56.5184 & 0.1413 & 1.487 & -- & 50.0945 & 0.1252 & 0.5686 & -- & 39.0766 & 0.0977 & 0.1469 & -- & 20 & 0.05\\
 \hline\hline
\multicolumn{17}{|c|}{Case 2: $p=1$}\\
\hline
&\multicolumn{4}{|c|}{$n=0.4$}&\multicolumn{4}{|c|}{$n=0.6$}&\multicolumn{4}{|c|}{$n=0.8$}&\multicolumn{4}{|c|}{$n=1$}\\
\hline
$p$  & $t_{Inf_1}$ & $t_{Inf_2}$ & $x(t_{Inf_1})$ & $\pi_x$ & $t_{Inf_1}$ & $t_{Inf_2}$ & $x(t_{Inf_1})$ & $\pi_x$ & $t_{Inf_1}$ & $t_{Inf_2}$ & $x(t_{Inf_1})$ & $\pi_x$ & $t_{Inf_1}$ & $t_{Inf_2}$ & $x(t_{Inf_1})$ & $\pi_x$ \\
\hline
1 &   16.0872         & -- & 172.4805 &   0.4312   & 13.5929 & -- & 182.7511   & 0.4569   & 12.52   &-- &  191.8533 & 0.4796   & 11.9779  & -- & 200  & 0.5\\
\hline\hline
\multicolumn{17}{|c|}{Case 3a: $0<p<1$, $m=\frac{1}{1-p}$ even}\\
\hline
&\multicolumn{4}{|c|}{$n=0.4$}&\multicolumn{4}{|c|}{$n=0.6$}&\multicolumn{4}{|c|}{$n=0.8$}&\multicolumn{4}{|c|}{$n=1$}\\
\hline
$p$  & $t_{Inf_1}$ & $t_{Inf_2}$ & $x(t_{Inf_1})$ & $\pi_x$ & $t_{Inf_1}$ & $t_{Inf_2}$ & $x(t_{Inf_1})$ & $\pi_x$ & $t_{Inf_1}$ & $t_{Inf_2}$ & $x(t_{Inf_1})$ & $\pi_x$ & $t_{Inf_1}$ & $t_{Inf_2}$ & $x(t_{Inf_1})$ & $\pi_x$ \\
\hline
 $1/2$ & 27.5175 & 35.6825 & 272.0777 & 0.6802 & 36.469 & 42.8742 & 282.9864 & 0.7075 & 52.0274 & 57.3726 & 292.1658 & 0.7304 & 77.5905 & 82.2093 & 299.9976 & 0.75\\
$3/4$ & 21.0996 & 50.0059 & 218.8878 & 0.5472 & 21.98 & 43.071 & 230.687 & 0.5767 & 24.6667 & 41.4846 & 240.9605 & 0.6024 & 28.7138 &  42.7956 & 249.9994 & 0.625\\
$5/6$ & 19.3102 & 68.7366 & 202.6804 & 0.5067 & 18.7055 & 53.7798 & 214.214 & 0.5355 & 19.5685 & 47.0293 & 224.345 & 0.5609 & 21.2138 & 43.905 & 233.3333 & 0.5833\\
$7/8$ &18.4638 & 88.1987  & 194.8557 & 0.4871 & 17.268 & 66.0313 & 206.1615 & 0.5154 & 17.4721 & 55.2748 & 216.1276 & 0.5403 & 18.3201 & 49.3305 & 224.9994 & 0.5625\\
\hline\hline
\multicolumn{17}{|c|}{Case 3b: $0<p<1$, $m=\frac{1}{1-p}$ even}\\
\hline
&\multicolumn{4}{|c|}{$n=0.4$}&\multicolumn{4}{|c|}{$n=0.6$}&\multicolumn{4}{|c|}{$n=0.8$}&\multicolumn{4}{|c|}{$n=1$}\\
\hline
$p$  & $t_{Inf_1}$ & $t_{Inf_2}$ & $x(t_{Inf_1})$ & $\pi_x$ & $t_{Inf_1}$ & $t_{Inf_2}$ & $x(t_{Inf_1})$ & $\pi_x$ & $t_{Inf_1}$ & $t_{Inf_2}$ & $x(t_{Inf_1})$ & $\pi_x$ & $t_{Inf_1}$ & $t_{Inf_2}$ & $x(t_{Inf_1})$ & $\pi_x$\\
\hline
$2/3$ & 23.0406 & 32.301 & 235.849 &  0.5896 & 25.9047  & 2.8383 & 247.6445 & 0.6191 & 31.3393 &  36.9606 & 257.8085 &  0.6445 & 39.4093 & 44.1715 & 266.6667 & 0.6667\\
$4/5$ & 20.0096 &  39.6109 &  209.0736 &  0.5227 & 19.9468 & 34.0155 & 220.7446 & 0.5519 & 21.4492 & 32.5474 & 230.9627 &  0.5774 & 23.9069 &  33.128 &  240 &  0.6\\
$6/7$ & 18.8229 &  48.6223 &  198.1865 & 0.4955 & 17.8694 & 38.8403 &  209.5974 & 0.524 & 18.3383 &  34.6657 & 219.6418 & 0.5491 & 19.5016 & 32.9379 & 228.5714 & 0.5714\\
$8/9$ & 18.1882 &  58.1104 &  192.2885  & 0.4807 & 16.8149 & 44.5911 & 203.5048  & 0.5088 & 16.8295 & 38.2891 & 213.4022 &  0.5335 & 17.457 & 35.0162 & 222.2222 & 0.5556\\
\hline\hline
\multicolumn{17}{|c|}{Case 3c: $0<p<1$, $m=\frac{1}{1-p} \notin {\mathbb N}$}\\
\hline
&\multicolumn{4}{|c|}{$n=0.4$}&\multicolumn{4}{|c|}{$n=0.6$}&\multicolumn{4}{|c|}{$n=0.8$}&\multicolumn{4}{|c|}{$n=1$}\\
\hline
$p$  & $t_{Inf_1}$ & $t_{Inf_2}$ & $x(t_{Inf_1})$ & $\pi_x$ & $t_{Inf_1}$ & $t_{Inf_2}$ & $x(t_{Inf_1})$ & $\pi_x$ & $t_{Inf_1}$ & $t_{Inf_2}$ & $x(t_{Inf_1})$ & $\pi_x$ & $t_{Inf_1}$ & $t_{Inf_2}$ & $x(t_{Inf_1})$ & $\pi_x$ \\
\hline
$1/6$ & 29.4776 & -- & 354.0681 & 0.8852 & 47.4915 & -- & 359.2075 & 0.898 & 83.1119 & -- & 363.3125 & 0.9083 & 153.539 & -- & 366.6667 & 0.9167\\
$1/5$ & 27.0472 & -- & 345.2826 & 0.8632 & 40.7017 & -- & 351.2607 & 0.8782 & 66.6889 & -- & 356.0608 & 0.8902 & 115.4096 &  -- & 360 &  0.9\\
$1/4$ & 24.075 & -- & 332.352 & 0.8309 & 32.8761 & -- & 339.4741 &  0.8487 & 49.0508 & -- & 345.2386 &  0.8631 & 77.3561 & -- & 350 &  0.875\\
$1/3$ & 20.539 & -- & 311.4553 & 0.7786 & 24.1889  & -- & 320.1892 &  0.8005 & 31.2983  & -- & 327.3521 & 0.8184 & 42.8587 & -- & 333.3333 & 0.8333\\
\hline\hline
\end{tabular}
}
\end{sideways}
\end{center}
\caption{{\scriptsize Values of $t_{Inf_1}$, $t_{Inf_2}$, $x(t_{Inf_1})$ and $\pi_x$ are listed for $n=0.4, 0.6, 0.8$ and $1$ and several choices of $p$.}}
\label{TABELLA}
\end{table}
\newpage
%
\section{Conclusions and future works}
In this work we have considered a growth model described by means of an ODE, able to include the most used growth curves, such as Gompertzian, Logistic, Bertalanffy-Richards and Malthus. The analytical solution $x(t)$ has been provided and a study of this solution has been made. The considered growth curve is characterized by the presence of several shape parameters and we focused on two of them, $p$ and $n$, with $0<p<1+1/n$ in \eqref{ODE} and in \eqref{sol1}. In particular, by considering three cases for the parameter $p$, different behaviors of the growth curve have been highlight in terms of domain, monotonicity, convexity and asymptotic behavior. The three considered cases are:
\begin{enumerate}
      \item $1<p<1+\frac{1}{n}$;
		\item $p=1$;
 		\item $0<p<1$.
\end{enumerate}
In the Cases 1 and 2 the curve is defined in $[t_0,\infty)$, it is monotonically strictly increasing, showing a sigmoidal behavior and it tends to its carrying capacity $k$ for $t\to \infty$. Further, there exists a unique inflection point at $\big(t_{Inf}, x(t_{Inf})\big)$;  the values of $t_{Inf}$ and $x(t_{Inf})$ decrease as $p$ increases, whereas for increasing values of $n$, $t_{Inf}$ decreases while $x(t_{Inf})$ increases. \par
More interesting is the Case 3, in which $x(t)$ exhibits different behaviors depending on the value of the parameter $p$. In particular, we considered three subcases of the Case 3:
\begin{enumerate}
      \item[3a.] $\dfrac{1}{1-p}=m\in\mathbb{N}$, with $m$ even;
      \item[3b.] $\dfrac{1}{1-p}=m\in\mathbb{N}$, with $m$ odd;
	  \item[3c.] $\dfrac{1}{1-p}=m\notin\mathbb{N}$.
\end{enumerate}
In the Case 3a the curve $x(t)$ is defined in $[t_0,\infty)$ and it has a bell shape, so it vanishes for $t\to\infty$. Near the maximum, corresponding to $k$, a plateau is highlighted whose amplitude increases as $p$ increases.
In the case $p=\frac{1}{2}$ it seems that the role of $n$ is to  translate ahead the curve as $n$ increases.
In the Case 3b the curve $x(t)$ is defined in a finite time interval and it explodes in a finite time.  Further, $x(t)$ presents a plateau near the parameter $k$, which width increases as $p$ increases, then $x(t)$ presents an exponential behavior.
Finally, in the Case 3c the curve $x(t)$ is defined in a finite time interval $[t_0,t_2)$, whose amplitude increases as $n$ increases and decreases for increasing $p$; moreover, $\lim_{t\to t_2} x(t)=k$, so the population reaches its carrying capacity in a finite time interval.\par
The study carried out in this paper lays the foundations for a generalization to growth in a stochastic environment, capable of putting in a single framework even very different behaviors of the growth phenomenon and thus providing for them a single mathematical tool for probabilistic-statistical analysis.
\end{document}